\documentclass[onecolumn,superscriptaddress,nofootinbib,notitlepage]{revtex4-2}

\usepackage[utf8]{inputenc}
\usepackage{amsfonts}
\usepackage{dcolumn}
\usepackage{float}
\usepackage{bm}
\usepackage{graphicx}
\usepackage{caption}
\usepackage{subcaption}
\usepackage{appendix}
\usepackage{amssymb,amsmath}
\usepackage{comment}
\usepackage[hidelinks]{hyperref}
\usepackage{ushort}
\usepackage[normalem]{ulem}
\usepackage{xcolor}
\usepackage{soul}
\hypersetup{colorlinks=true, linkcolor=blue}
\usepackage{ragged2e}
\DeclareCaptionJustification{justified}{\justifying}
\begin{document}

\title{ Born--Infeld Electrogravity and Dyonic Black Holes}
\author{Guadalupe Ahumada Acu\~{n}a \footnote{Corresponding author.}}
\email{gi.ahumadaacuna@df.uba.ar} \affiliation{Universidad de Buenos
Aires, Facultad de Ciencias Exactas y Naturales, Departamento de
F\'{\i}sica. Buenos Aires, Argentina.} \affiliation{CONICET - Universidad
de Buenos Aires, Instituto de F\'{\i}sica de Buenos Aires (IFIBA). Buenos
Aires, Argentina.}

\author{Cecilia Bejarano} \email{cbejarano@iafe.uba.ar}
\affiliation{Instituto de Astronom\'{\i}a y F\'{\i}sica del Espacio (IAFE,
CONICET-UBA). Buenos Aires, Argentina.}

\author{Rafael Ferraro}
\email{ferraro@iafe.uba.ar} \affiliation{Instituto de Astronom\'{\i}a y
F\'{\i}sica del Espacio (IAFE, CONICET-UBA). Buenos Aires, Argentina.}

\begin{abstract}

Born--Infeld electrogravity is defined through a Lagrangian that
couples gravity and electromagnetism within a single determinantal
structure. The field equations are derived in Palatini's formalism,
where the metric, connection, and vector potential are varied
independently in the action. As a result, the gravitational sector
reduces to Einstein's equations with a torsion-free,
metric-compatible connection. The electrodynamic sector, in turn,
admits two equivalent interpretations or \textit{pictures}: it can
be seen either as a standard Born--Infeld electrodynamics in an
effective background geometry, or as an \textit{anomalous}
Born--Infeld electrodynamics in the physical metric. We illustrate
the dynamics by analyzing the horizon structure, the extremality
conditions, and the thermodynamics of spherically symmetric dyonic solutions. 
\end{abstract}

\maketitle
\section{Introduction}
Born--Infeld theory (BI) has its origins in the context of nonlinear
electrodynamics. Proposed by Max Born and Leopold Infeld in 1934
\cite{born-33, borninfeld-34-Nature, born-34, borninfeld-34}, this
theory emerged as an extension of classical Maxwell
electromagnetism, with the aim of removing the divergences in the
self-energy of the  point-like charged particle by the following
action
\begin{equation}\label{eq:bi}
S \propto \int d^4x \sqrt{-\det \left( g_{\mu\nu} + b^{-1} F_{\mu\nu}
\right)} - \lambda \sqrt{-\det (g_{\mu\nu})},
\end{equation}
where $g_{\mu\nu}$ is the metric tensor, $F_{\mu\nu}$ is the
electromagnetic field tensor, $b$ is a new fundamental constant
which sets  the order of magnitude where the nonlinear effects
become relevant, and $\lambda$ controls the asymptotic behavior of
vacuum solutions, with $\lambda = 1$ corresponding to Minkowski
spacetime. In the weak field limit, the BI theory reduces to the
Maxwell action plus small corrections, while for the strong regime
the field equations deviate significantly from Maxwell's
electromagnetism.  This well known capacity to cure the
singularities motivated the so-called BI gravity theories, where the
gravitational action is built by analogy with the BI structure (Eq.
\eqref{eq:bi}) through the Ricci tensor $R_{\mu\nu}$
\begin{equation}\label{eq:big}
S \propto \int d^4x \sqrt{-\det \left( g_{\mu\nu} + \epsilon R_{\mu\nu}
\right)} - \lambda \sqrt{-\det (g_{\mu\nu})}.
\end{equation}
This action was studied in the standard metric formalism
\cite{deser,freigen, freigen4D} as well as in the Palatini
(metric-affine) approach \cite{vollick04} with the intention of
regularizing the spacetime singularities inside black holes and also
in the initial singularity at the beginning of the Universe in the
Big Bang  model \cite{banados}. For a recent review on BI inspired
modifications of gravity, see \cite{review_beltran}.

Many modified gravity theories inspired by the BI action introduce
matter fields following the standard prescription of minimal
coupling in metric theories of gravity (MTG) \cite{MTG}. This
approach ensures consistency with Einstein's equivalence principle
\cite{Einstein_equivalence}, and splits the total action into a
gravitational part and a matter part: $S_{\text{total}} =
S_{\text{gravity}} + S_{\text{matter}}$. The matter sector is then
constructed from the Minkowskian metric promoted to a curved
spacetime metric $g_{\mu\nu}$, making use of covariant derivatives
$\nabla_\mu$, and some minimal coupling prescription (which is not
always free of ambiguities \cite{ambiguedades}). Explicitly, one
writes $S_{\text{matter}} = S[g_{\mu\nu}, \psi_m, \nabla \psi_m]$,
where $\psi_m$ generically denotes the matter fields. Perhaps for
this reason, attempts to unify the matter and gravity sectors within
a single determinantal action have received relatively little
attention. To the best of our knowledge, the notable exception is
the class of models proposed by Vollick in \cite{Vollick05}, who
introduced an action of the form
\begin{equation}\label{eq:vollick}
S \propto \int d^4x \sqrt{-\det \left( g_{\mu\nu} + \epsilon
R_{\mu\nu}(\Gamma) + \beta M_{\mu\nu} \right)} - \lambda \sqrt{-\det
(g_{\mu\nu})},
\end{equation}
where $M_{\mu\nu}$ denotes a tensor constructed from matter fields,
and $\epsilon$ and $\beta$ are universal constants of the theory. For massive scalar fields one may take
$M_{\mu\nu} = \partial_\mu \phi \partial_\nu \phi$, while for
electromagnetic fields $M_{\mu\nu} = F_{\mu\nu}$. A construction of
this type does not fit into the usual separation of matter and
gravity in theories based on the principle of minimal coupling,
which suggests that such models should eventually conflict with the
experimental evidence supporting the equivalence principle.
Nevertheless, Vollick showed that for $M_{\mu\nu} = \partial_\mu
\phi \partial_\nu \phi$, the field equations coincide with those of
General Relativity minimally coupled to a free scalar field. For electromagnetic and
fermionic fields, at the first non-trivial order in perturbations,
one recovers the usual Einstein-Maxwell and Einstein-Dirac systems,
indicating that such theories effectively reduce to an MTG
representation at that order.

In this work we follow Vollick's pioneering proposal of Eq.~\eqref{eq:vollick}, with $M_{\mu\nu}=F_{\mu\nu}$, recently revisited by Afonso et al \cite{afonso_et_al}. To derive the field equations we vary the action \textit{\`{a} la} Palatini, where the connection and the metric are treated as independent variables. In the standard metric formalism, instead, the connection is given in terms of the metric (namely, the Levi-Civita connection), so that the two objects are not independent.  It is well known that Palatini and metric formalisms lead to the same dynamics in General Relativity, Lovelock theories \cite{Exirifard} and other particular cases where the dynamical equations end up being second order because the higher derivatives go to boundary terms in the metric formalism. But in general, the metric and Palatini formalisms are two ways of varying the action that lead to different dynamics. In many modified gravity theories, varying the action within the metric formalism leads to fourth-order dynamical equations for the metric, potentially introducing extra degrees of freedom and instabilities such as Ostrogradski ghosts (see \cite{Sotiriou} and references therein). By contrast, varying the action \textit{\`a la} Palatini \cite{Palatini} typically leads to second-order field equations for the metric, thus avoiding the higher-derivative pathologies that appear in the metric approach. However, the Palatini approach is not entirely free of problems. Ghosts may reappear in scalar-tensor theories with matter coupled to the connection (however, see \cite{Galtsov,Aoki}), and the Cauchy problem may be ill-possed \cite{Faraoni} (however, see \cite{Palatini}). Nevertheless, these concerns are not relevant for the BI gravito-electromagnetic Lagrangian considered here, since the Palatini variation leads to Einstein equations whose electromagnetic source is a metric energy-momentum tensor. Suggestively, this resembles the correspondence between modified gravity and
General Relativity plus different matter fields  \cite{rbg1,rbg2}.

The article is organized as follows. Section~\ref{sec.BI electrogravity} introduces the gravito-electromagnetic Lagrangian with non-minimal coupling (Eq.~\eqref{eq:vollick}), and we obtain the field equations in the Palatini formalism. Section~\ref{sec:Dynamics} exploits the structure of BI-like Lagrangians to obtain the dynamics. Section~\ref{sec:eqv_pr} shows that, despite appearances, the system evolves consistently with an MTG theory. 
We show that the gravitodynamics reduces to Einstein's equations with a torsion-free, metric-compatible connection, while the electrodynamics admits two
equivalent interpretations or pictures: (i) a standard BI
electrodynamics in an effective background geometry, or (ii) an
anomalous BI electrodynamics in the physical metric. Section~\ref{cap:dionic_bh} presents the results concerning the dyonic BH, which carries both electric and magnetic charges. We study the horizon structure, the extremality conditions and the thermodynamics. In Section~\ref{sec:concl}, we present the conclusions of the work.

\section{Born--Infeld electrogravity} \label{sec.BI electrogravity}

\subsection{Born--Infeld electrodynamics}
Born--Infeld Lagrangian was initially introduced as an idea to avoid
the divergence of the electrostatic field of a point-like charge
\cite{born-33,borninfeld-34-Nature,born-34}. In its most primitive
version, it simply mimicked the Lagrangian of relativistic
mechanics, since its electrostatic form was
\begin{equation}
    L\propto\sqrt{1-b^{-2}\mathbf{E}^2}\ .
\end{equation}
This was a successful attempt, in the sense that the electric field
$\mathbf{E}$ of a point-like charge turned out to be bounded above
by the new BI constant $b$, in the same way that $c$ is an
upper limit for speed in relativistic mechanics. Born and Infeld
soon realized that this Lagrangian was the Minkowskian version of a
more general one, whose geometric formulation in terms of the
electromagnetic field $F_{\mu\nu}=\partial_\mu A_\nu-\partial_\nu
A_\mu$ is given by the density \cite{borninfeld-34}
\begin{equation}
    L _{BI}[A]=-\frac{b^2}{4\pi}\; \left(\sqrt{|\det(g_{\mu\nu}+b^{-1}F_{\mu\nu})|}-\sqrt{|\det(g_{\mu\nu})|}\right)\ .
\end{equation}
The last term in $L _{BI}$ is necessary to ensure the correct Maxwellian
limit when the field is weak compared to $b$; namely, it guarantees
that the energy-momentum goes to zero when the field goes to zero.
The weak-field limit can be easily checked by writing
$\det(g_{\mu\nu}+b^{-1}F_{\mu\nu})$ in terms of the scalar $S\equiv
F^{\mu\nu}F_{\mu\nu}/4$ and the pseudoscalar $P\equiv
{^\ast}F^{\mu\nu} F_{\mu\nu}/4$,  where ${^\ast}F^{\mu\nu}$ is the
dual of
$F_{\mu\nu}$.\footnote{$\det(g_{\mu\nu}+b^{-1}F_{\mu\nu})=\det(g_{\mu\nu})(1
+ b^{-2} 2S -b^{-4} P^2)$, as shown in Appendix
\ref{apendice:desarrollo_q}. In Minkowski spacetime it is
$2S=\mathbf{B}^2-\mathbf{E}^2 $ and $P=\mathbf{E}\cdot\mathbf{B}$.}

BI electrodynamics can be summarized through its dynamical
equation and energy-momentum tensor,
\cite{Tmunu_BI2,Tmunu_BI3,Tmunu_BI4,desarrollo_Tmunu_BI}
\begin{equation} \label{eq born infeld}
\partial_\mu \left(\sqrt{-g} \;\mathcal{F}^{\mu\nu}\right) = 0\ ,
\end{equation}
\begin{equation} \label{T munu BI}
T^{\mu \nu}_{BI} \equiv -\frac{2}{\sqrt{-g}}\;\frac{\delta
L_{BI}}{\delta g_{\mu \nu}}= \frac{1}{4\pi} \left[{\mathcal
F}^{\mu\rho }\, F^{\;\nu}_{\rho}-b^2\;  g^{\mu \nu} \left(1 -
\sqrt{1 + b^{-2} 2S - b^{-4} P^2} \right)\right]\ ,
\end{equation}
(metric signature $+---$; Gaussian units) where \(g \equiv
\det(g_{\mu\nu})\), and \(\mathcal{F}^{\mu\nu}\) is
\begin{equation} \label{cursive F bornInfeld}
\mathcal{F}^{\mu\nu} \equiv \frac{ F^{\mu\nu} - b^{-2} P \
{}^{\ast}F{^{\mu\nu}}} {\sqrt{1 + b^{-2} 2S - b^{-4} P^2}}\ .
\end{equation}

\subsection{Born--Infeld electrogravity}
Inspired by BI nonlinear electrodynamics, and motivated by
its capabilities of smoothing singularities, we study an
electrogravity theory governed by the Lagrangian
\cite{Vollick05,afonso_et_al} 
\begin{equation} \label{eq accion}
    L \propto  \sqrt{-q} - \lambda \sqrt{-g} \ ,
\end{equation}
where \(q \equiv \det(q_{\mu\nu})\) is the determinant of an
auxiliary tensor, defined as
\begin{equation}\label{eq q}
    q_{\mu\nu} \equiv g_{\mu\nu} + \epsilon R_{(\mu\nu)}(\Gamma) + \beta F_{\mu\nu}(A).
\end{equation}
\(F_{\mu\nu}(A)=\partial_\mu A_\nu-\partial_\nu A_\mu\) is the
electromagnetic field tensor, and \(R_{(\mu\nu)}(\Gamma)\) is the
symmetric part of the Ricci tensor, constructed from a
metric-independent affine connection \( \Gamma^\lambda_{\mu\nu}\),
given by \footnote{We use the convention $V^\mu_{\ ;\nu}=V^\mu_{\
,\nu}+\Gamma^\mu_{\rho\nu}V^\rho$.}
\begin{equation} \label{eq ricci}
    R_{\mu\nu}(\Gamma) \equiv \partial_\alpha \Gamma^\alpha_{\mu\nu} - \partial_\nu \Gamma^\alpha_{\mu\alpha} + \Gamma^\beta_{\mu\nu} \Gamma^\alpha_{\beta\alpha} -\Gamma^\beta_{\mu\alpha} \Gamma^\alpha_{\beta\nu} .
\end{equation}
The parameters \(\epsilon\) and \(\beta\) are fundamental constants
of the theory, with dimensions of squared length and inverse field
strength, respectively, that satisfy $\epsilon^{-1}\beta^2\propto G$, where $G$ is the Newton constant.  Finally, \(\lambda \neq 1\) indicates the presence of a
cosmological constant \(\Lambda = \epsilon^{-1}(1-\lambda)\), as
will be shown.

\subsection{Dynamical equations \textit{\`{a} la} Palatini}\label{Palatini}

In general, the variation of the determinant of a matrix $ M$ is
\begin{equation}
\delta\;\ln|\det M|=\operatorname{Tr}({\bar{M}}\cdot \delta M)\ ,
\end{equation}
where \(\bar{M}\) denotes the inverse of \({M}\). Thus, the
variation of the Lagrangian~(\ref{eq accion}) yields
\begin{equation}
    \delta L \propto  \sqrt{-q}\, {\bar q}^{\nu\mu}\,  \delta q_{\mu\nu} - \lambda \sqrt{-g}\,  g^{\mu\nu} \, \delta g_{\mu\nu}\ ,
\end{equation}
where \( \bar{q}^{\mu\nu} \) is the inverse of \( q_{\mu\nu} \),
\begin{equation}
 q_{\mu\alpha}\, {\bar q}^{\alpha\nu}= \delta_\mu^\nu\ ,
\end{equation}
and
\begin{equation}\label{deltaqmunu}
     \delta q_{\mu\nu} = \delta g_{\mu\nu} + \epsilon\, \delta R_{(\mu\nu)}(\Gamma) + \beta\, \delta F_{\mu\nu}({A})\ .
\end{equation}

Since we adopt the Palatini approach, the variational procedure  treats \({\bf g}\), \({\bf \Gamma}\), and \({\bf A}\) as independent dynamical variables. Therefore, the
variation of the Lagrangian with respect to each of them leads to
three equations that govern the coupled dynamics of the metric, the
affine connection, and the electromagnetic field.

The equation resulting from the variation with respect to the
metric,
\begin{equation} \label{eq grav}
   \sqrt{-q}\,\bar{q}^{(\mu\nu)} - \lambda \sqrt{-g}\, g^{\mu\nu} = 0\,,
\end{equation}
is not strictly a dynamical equation since it does not contain
second-order time derivatives; actually it is a constraint equation,
as it relates the canonical variables and their momenta. This is a
consequence of the absence of metric derivatives in the Lagrangian.
Instead the Lagrangian contains derivatives of \({\bf \Gamma}\) and
\({\bf A}\); so the dynamics will follow from varying it with
respect to these variables.

The variation with respect to \({\bf A}\) leads to a dynamical
equation for the electromagnetic field in a given geometric
background. Since \(\delta F_{\mu\nu} = 2\,\partial_{[\mu} \delta
A_{\nu]}\), after integrating by parts one obtains
\begin{equation}\label{eq e.m.}
\partial_{\mu}\!\left( \sqrt{-q}\, \bar{q}^{[\mu\nu]} \right) = 0\,.
\end{equation}
Thus, the symmetric and antisymmetric parts of the tensor density
\(\sqrt{-q}\;\bar{q}^{\mu\nu}\) separately enter the constraint
equation (\ref{eq grav}) and the dynamical equation (\ref{eq e.m.}),
as a nice consequence of the symmetric and antisymmetric characters
of \(\delta g_{\mu\nu}\) and \(\delta F_{\mu\nu}\) respectively.

On the other hand, the variation of the Ricci tensor in
Eq.~(\ref{eq ricci}) with respect to the affine connection \({\bf
\Gamma}\) can be written as
\begin{equation}\label{varricci}
\delta R_{\mu\nu} = \nabla^\Gamma_\alpha (\delta
\Gamma^\alpha_{\mu\nu}) - \nabla^\Gamma_\nu (\delta
\Gamma^\alpha_{\mu\alpha}) + T^\theta_{\ \lambda\nu}\, \delta
\Gamma^\lambda_{\mu\theta}\ ,
\end{equation}
where $ T^\theta_{\ \alpha\beta} \equiv 2
\Gamma^\theta_{[\beta\alpha]}$ is the torsion tensor. In this
expression for \(\delta R_{\mu\nu}\) we take advantage of the fact
that the difference \(\delta{\bf \Gamma}\) between two affine
connections is a tensor (even though \({\bf \Gamma}\) is not).

The symmetric character attributed to \(R_{\mu\nu}\) in Eq.~(\ref{eq q})
forces the variation \(\delta{\bf \Gamma}\) to ensure that \(\delta
R_{[\mu\nu]} = 0\); therefore, the variation \(\delta{\bf \Gamma}\)
is not arbitrary.  In Eq.~(\ref{varricci}), the term containing \(\delta \Gamma^\alpha_{\mu\alpha}\) effectively contributes to \(\delta R_{[\mu\nu]}\); but it is a \textit{projective}-type contribution~\cite{projective2,
invarianciaproyectiva}, which is degenerate with a similar contribution of \(\delta A_\mu\) to the
variation of the Lagrangian (see Appendix
\ref{app:Projective}), and can be absorbed into its electromagnetic counterpart. Hence, the condition \(\delta R_{[\mu\nu]} =
0\) requires only that the torsion be set to zero. In this way, the third term in Eq.~(\ref{varricci}) does not appear, and 
\(\delta \Gamma^\alpha_{[\mu\nu]}\) in the first term vanishes because the torsion has been fixed.\footnote{The imposition of zero torsion within the Palatini formalism should not surprise. The proof of the equivalence between the metric and Palatini formalisms in General Relativity and Lovelock theories presupposes that torsion is zero or \(\Gamma\) is a metric connection.}  
Thus, Eq.~(\ref{varricci}) becomes
\begin{equation}
 \delta R_{\mu\nu}=\nabla^\Gamma_\alpha (\delta \Gamma^\alpha_{(\mu\nu)})\ ,
\end{equation}
and the corresponding variation of the Lagrangian yields the dynamical equation \footnote{As follows from the integration by
parts that leads to this equation,
\(\nabla^\Gamma_\alpha\left(\sqrt{-q}\,\xi^{\nu\mu}\right)\) denotes
the tensor density \(\sqrt{-q}\,\nabla^\Gamma_\alpha \xi^{\nu\mu} +
\xi^{\nu\mu}\,\partial_\alpha \sqrt{-q}\).}
\begin{equation} \label{eq:conexion2}
\nabla^\Gamma_\alpha \left(\sqrt{-q}\, {\bar q}^{(\nu\mu)}\right) =
0 \ .
\end{equation}
However, the constraint~(\ref{eq grav}) turns this equation into the
condition for the connection being metric-compatible. Combined with
the torsion-free condition imposed above, this uniquely determines
the connection to be the Levi-Civita connection. Thus, Eqs.~(\ref{eq
grav}) become dynamic, since it now contains second derivatives of
the metric. Equations~(\ref{eq grav}) and~(\ref{eq e.m.}) govern the
dynamics of the metric and electromagnetic fields. To make practical
use of this system of equations, one must either explicitly
construct the tensor \(\bar{q}^{\mu\nu}\), or find a shortcut to
avoid that task. In this sense, we may notice that from the
electrodynamical point of view, the Lagrangian (\ref{eq
accion})--(\ref{eq q}) corresponds to a BI electrodynamics
in a geometry where the symmetric tensor
\begin{equation}\label{supermetrica}
    \mathcal{G}_{\mu \nu} \equiv g_{\mu \nu} + \epsilon R_{(\mu \nu)}
\end{equation}
plays the role of the background metric. This means that
Eq.~(\ref{eq e.m.}) must correspond to a BI equation in the
``metric'' $\mathcal{G}_{\mu \nu}$. On the other hand, the variation
of the Lagrangian (\ref{eq accion})--(\ref{eq q}) with respect to
the metric $g_{\mu\nu}$ should be related to the energy-momentum
tensor of the BI field in the background geometry
$\mathcal{G}_{\mu \nu}$. Since BI electrodynamics in
arbitrary geometric backgrounds is well known, these elements could
help to gain insight into several aspects of BI
electrogravity, before undertaking the task of building the tensor
\(\bar{q}^{\mu\nu}\).

\section{Dynamics of Born--Infeld electrogravity} \label{sec:Dynamics}
\subsection{The inverse of $q_{\mu\nu}$} \label{sec:Inverse q}
Any matrix $q$ can be decomposed into the sum of its symmetric and
antisymmetric parts, $\mathcal{G}$ and $F$,
\begin{equation}
q=\mathcal{G}+F= \mathcal{G}(I+\bar{\mathcal{G}}\ F)\equiv
\mathcal{G}(I+\breve{F})\ ,
\end{equation}
$\bar{\mathcal{G}}$ being the inverse of matrix $\mathcal{G}$. Note that we will use the tilde `` $\breve{}$ '' for any magnitude built with $\mathcal{G}$ or its inverse.  The inverse of $I+\breve{F}$ can be represented as a series of powers of $\breve{F}$ \footnote{$I=(I+\breve{F})~(I-\breve{F}+\breve{F}\breve{F}-\breve{F}\breve{F}\breve{F}%
+...)$, where each added term to the sum corrects the result for the
previous sum. Certainly, this nothing more than Taylor's series for
$f(x)=(1+x)^{-1}$.}
\begin{equation}
(I+\breve{F})^{-1}=\sum_{n=0}^{\infty }(-\breve{F} )^{n}~.
\end{equation}
But in four dimensions there exists another antisymmetric matrix
which helps us to express $(I+\breve{F})^{-1}$ in terms of $F$
without resorting to a series: the dual matrix of $F$ whose
components are \footnote{The ``metric'' role of $\mathcal{G}$ also
implies that  $^{\ast }\breve{F}_{\mu \nu }=
\frac{\sqrt{-\mathcal{G}} }{2}~\epsilon _{\lambda \rho \mu \nu
}~{\breve F}^{\lambda \rho }=\frac{1 }{2}~\breve{\epsilon} _{\lambda
\rho \mu \nu }~{\breve F}^{\lambda \rho }$, where ${\breve
F}^{\lambda \rho
}=\bar{\mathcal{G}}^{\lambda\mu}F_{\mu\nu}\bar{\mathcal{G}}^{\nu\rho}$.
Besides, ${^\ast}{^\ast}\equiv -1$. }
\begin{equation}
^{\ast }\breve{F}^{\mu \nu }\equiv \frac{1}{2\,
\sqrt{-\mathcal{G}}}~\epsilon ^{\lambda \rho \mu \nu }~F_{\lambda
\rho }=\frac{1}{2}\ \breve{\varepsilon}^{\lambda \rho \mu \nu
}~F_{\lambda \rho }~,
\end{equation}
where $\epsilon ^{\lambda \rho \mu \nu }$ is the Levi-Civita symbol
($-\epsilon ^{0123}=1=\epsilon_{0123}$), and $\breve{\varepsilon
}^{\lambda \rho \mu \nu }$ is the Levi-Civita tensor associated with
the ``metric'' $\mathcal{G}$.

We will operate with the matrices $\breve{F}$ and $^{\ast
}\breve{F}$ of components \footnote{The product of matrices is the
contraction of inner indices: $\bar{q}\, q=I$ means $\bar{q}^{\mu
\lambda }q_{\lambda \nu }=\delta^\mu_\nu$. }
\begin{equation}
\breve{F}_{~\nu }^{\mu }=\bar{\mathcal{G}}^{\mu \lambda }F_{\lambda
\nu }\ ,\ \ \ \ \ \ ^{\ast }\breve{F} _{~\nu }^{\mu }= {^{\ast
}\breve {F}} ^{\mu \lambda }~ \mathcal{G}_{\lambda \nu }~.
\end{equation}
$\breve{F}$ and $^{\ast }\breve{F}$ are the blocks to build two
matrices proportional to the identity:
\begin{eqnarray}
2\breve{S}~I &=&-\breve{F}\breve{F}+(^{\ast }{\breve F})~^{\ast
}\breve{F}~\,,
\label{relacion_S} \\
\breve{P}~I &=&-\breve{F}\,{}~^{\ast }\breve{F}=-(^{\ast }{\breve
F})\ \breve{F}\ ,  \label{relacion_P}
\end{eqnarray}

where the scalar $\breve{S}$ and the pseudoescalar $\breve{P}$ are
obtained by tracing the former equations:
\begin{align}
\breve{S}& =-\frac{1}{4}~\breve{F}_{~\lambda }^{\mu }\
\breve{F}_{~\mu }^{\lambda }=\frac{1}{4}~{}^{\ast
}\breve{F}_{~\lambda }^{\mu }\, ^{\ast }
\breve{F}_{~\mu }^{\lambda }~,  \label{eq:S} \\[0.04in]
\breve{P}& =-\frac{1}{4}{}~\breve{F}_{~\lambda }^{\mu }\, ^{\ast }\breve{F}%
_{~\mu }^{\lambda }=-\frac{1}{4}{}~{F}_{\mu\lambda }\, ^{\ast
}\breve{F}^{\lambda\mu }~.  \label{eq:relacion_P}
\end{align}
Now, let us compute
\begin{eqnarray}
(I+\breve{F})~(I-\breve{F}+{^{\ast }\breve F}\, ^{\ast
}\breve{F}+a~^{\ast } \breve{F})~
&=&I-\breve{F}~\breve{F}+{^{\ast }\breve F}~^{\ast }\breve{F}+a~^{\ast }%
\breve{F}+\breve{F}~^{\ast }\breve{F}~^{\ast }\breve{F}+a~\breve{F}~^{\ast }%
\breve{F} \notag\\
&=&I+2\breve{S}~I+a~^{\ast }\breve{F}-\breve{P}~I~^{\ast }\breve{F}-a~\breve{%
P}~I
\end{eqnarray}
By choosing $a$ equal to $\breve{P}$, one has the result that
\begin{equation}
(I+\breve{F})^{-1}=\frac{I-\breve{F}+{^{\ast }\breve F}~^{\ast }\breve{F}+%
\breve{P}~^{\ast }\breve{F}}{1+2\breve{S}-\breve{P}^{2}}
\end{equation}
So the inverse of $q=\mathcal{G}+\beta ~F=\mathcal{G}(I+\beta ~%
\breve{F})$ is
\begin{equation}
\bar{q}=\frac{I-\beta ~\breve{F}+\beta ^{2}~^{\ast }\breve{
F}~^{\ast }\breve{F}+\beta^3 ~\breve{P}~^{\ast }\breve{F}}{1+2\beta
^{2}~ \breve{S}-\beta ^{4}~\breve{P}^{2}}~\mathcal{\bar{G}}\ .
\end{equation}
By combining this expression for $\bar q$ with the value for
$\det{q}$ computed in Appendix \ref{apendice:desarrollo_q}
(Eq.~(\ref{detq})), one obtains
\begin{equation}\label{inverseq}
\sqrt{-q}\ \bar{q}^{\mu \nu
}=\sqrt{-\mathcal{G}}~\frac{\mathcal{\bar{G}}^{\mu \nu }-\beta
~\breve{F}^{\mu \nu }+\beta ^{2}~^{\ast }\breve{F}_{~\lambda }^{\mu
}~^{\ast }\breve{F} ^{\lambda \nu }+\beta ^{3}~\breve{P}~^{\ast
}\breve{F}^{\mu \nu }}{\sqrt{1+2\beta ^{2}~\breve{S}-\beta
^{4}~\breve{P}^{2}}}~.
\end{equation}
We end this preparatory subsection by noting that
\begin{equation}\label{dualdual}
  \sqrt{-\mathcal{G}}\  ^{\ast}\breve{F}^{\mu\lambda }=\frac{1}{2}\ \epsilon ^{\eta \rho \mu \lambda} ~F_{\eta \rho }=\sqrt{-g}\ ^{\ast }{F}^{\mu\lambda }~;
\end{equation}
 according to the relation (\ref{relacion_P}) this means that
\begin{equation}\label{relation_PP}
  \sqrt{-\mathcal{G}}\ \breve P= \sqrt{-g}\  P\ .
\end{equation}
On the other hand, by multiplying Eq.~(\ref{relacion_S}) with
$^\ast\breve  F$, and using Eq.~(\ref{relacion_P}), one gets
\begin{equation}\label{tripleF}
    2\breve{S}~^{\ast }\breve{F} =\breve{P}~\breve{F}+(^{\ast }{\breve F})~(^{\ast }{\breve F})~^{\ast }\breve{F}\ .
\end{equation}

\subsection{Electrodynamics}
In Eq.~(\ref{eq e.m.}) let us replace
$\sqrt{-q}\;\bar{q}^{[\mu\nu]}$ with the antisymmetric part of
Eq.~(\ref{inverseq}); then
\begin{equation} \label{eq em 0}
\boxed{
\partial_\mu \left( \sqrt{-\mathcal{G}} \ \frac{\breve{F}^{\mu\nu} - \beta^2\ \breve{P} \ ^\ast\breve{F}^{\mu\nu}}{\sqrt{1 +2\beta^2 \ \breve{S} - \beta^4\ \breve{P}^2}} \right) = 0\ .
}
\end{equation}
As anticipated, this equation governs the standard BI
electrodynamics in the geometric background described by the
``metric'' \(\mathcal{G}_{\mu\nu}\), since we recognize in
parentheses the respective tensor $\breve{\mathcal{F}}^{\mu\nu}$ as
introduced in Eq.~(\ref{cursive F bornInfeld}). Certainly, we do not
yet know \(\mathcal{G}_{\mu\nu}=g_{\mu\nu}+\epsilon R_{\mu\nu}\) and
its inverse \(\bar{\mathcal{G}}^{\mu\nu}\), since they are subject
to the gravitodynamic equations.

\subsection{Gravitodynamics}
Now, replace $\sqrt{-q}\;\bar{q}^{(\mu\nu)}$ in Eq.~(\ref{eq grav})
with the symmetric part of Eq.~(\ref{inverseq}); then
\begin{equation} \label{eq grav 1}
 \sqrt{-\mathcal{G}}~\frac{\mathcal{\bar{G}}^{\mu \nu }+\beta ^{2}~^{\ast }\breve{F}_{~\lambda }^{\mu }~^{\ast }\breve{F}
^{\lambda \nu }}{\sqrt{1+2\beta ^{2}~\breve{S}-\beta
^{4}~\breve{P}^{2}}}= \lambda \sqrt{-g}\, g^{\mu\nu}\ .
\end{equation}
After contracting this expression with \(\mathcal{G}_{\nu\rho}\),
and substituting
\begin{equation}
\sqrt{-\mathcal{G}} = \sqrt{-g} \;
 \sqrt{\det(\delta^{\mu}_{\nu} + \epsilon R^{\mu}_{\;\nu})}\ ,
\end{equation}
where \(R^{\mu}_{\;\nu} = g^{\mu \rho} R_{\rho \nu}\), the dynamical
equations for the geometry become
\begin{equation}  \label{eq grav 0}
\boxed{
  \lambda\,\frac{ \delta^{\mu}_{\nu}+\epsilon\, R^\mu_{\ \nu}}{\sqrt{\det( \delta^{\mu}_{\nu}+\epsilon\, R^\mu_{\ \nu})}}=\frac{\delta^{\mu }_{\nu }+\beta ^{2}~^{\ast }\breve{F}_{~\lambda }^{\mu }~^{\ast }\breve{F}
^{\lambda}_{ \ \nu }}{\sqrt{1+2\beta ^{2}~{S}-\beta
^{4}~\breve{P}^{2}}}\ .
}
\end{equation}
Ricci tensor both in Eqs.~(\ref{eq em 0}) and (\ref{eq grav 0}) is
written in terms of the Levi-Civita connection, as obtained from the
variational calculus in Section (\ref{Palatini}).

We remark that any geometry whose Ricci tensor is \(R^{\mu}_{\;\nu}
= -\Lambda \delta^{\mu}_{\nu}\) --in particular, de Sitter
geometry-- is a vacuum solution to Eq.~(\ref{eq grav 0}) for
$\lambda=1-\epsilon\Lambda$.\footnote{For $^{\ast
}\breve{F}^{\mu}_{\ \lambda}=0$ the equation reduces to $\lambda\,
\frac{(1-\epsilon \Lambda)}{\sqrt{(1-\epsilon \lambda)^4}}
\,\delta^\mu _\nu =\delta^\mu_\nu $.}

\section{Equivalence principle} \label{sec:eqv_pr}
Equations (\ref{eq em 0}) and (\ref{eq grav 0}) seem to imply that
the equivalence principle is violated by the action (\ref{eq
accion}). This impression is caused by the presence of the Ricci
tensor \(R_{\mu\nu}\) in Eq.~(\ref{eq em 0}) which governs the
dynamics of the electromagnetic field. In fact, the Ricci tensor
contributes to both the volume \(\sqrt{-\mathcal{G}}\) and the tensor
\(\breve{F}^{\mu\nu} = \bar{\mathcal{G}}^{\mu\lambda}
\bar{\mathcal{G}}^{\nu\rho} F_{\lambda\rho}\). In addition, the
source in the r.h.s of  Eq.~(\ref{eq grav 0}) is also contaminated
by the Ricci tensor. However, as shown in Ref. \cite{afonso_et_al},
the dynamical equations can be combined in such a way that both
undesired contributions of the Ricci tensor disappear.  We will use
the rest of this subsection to demonstrate this property.

Let us begin with Eq.~(\ref{eq grav 1}), which is the result of
varying the Lagrangian (\ref{eq accion}) with respect to the metric
\(g_{\mu\nu}\). This dynamical equation can be regarded as the
relation between the operations of raising indices with
$\bar{\mathcal{G}}^{\mu\nu}$ or $g^{\mu\nu}$.  Although the term
$^{\ast}F\ ^{\ast}F$ on the l.h.s of Eq.~(\ref{eq grav 1}) is
cumbersome, its presence can be easily handled with the help of
Eqs.~(\ref{relacion_P}) and (\ref{tripleF}), depending on whether
$^{\ast}F\ ^{\ast}F$  acts on $F$ or $^{\ast}F$. For instance, by
contracting Eq.~(\ref{eq grav 1}) with $F_{\nu\rho}$ and using
Eq.~(\ref{relacion_P}) one obtains
\begin{equation}\label{eq grav 2}
 \sqrt{-\mathcal{G}}~\frac{\breve{F}^\mu_{\ \rho }-\beta ^{2}\;\breve{P}\ ^{\ast }\breve{F}_{~\rho }^{\mu }}{\sqrt{1+2\beta
^{2}~\breve{S}-\beta ^{4}~\breve{P}^{2}}} = \lambda \sqrt{-g}\
F^\mu_{\ \rho} \ ,
\end{equation}
which expresses the relationship that dynamics establishes between
electrodynamic variables generated by the ``metric'' $\mathcal G$
and those associated with the metric $g$.  To further explore this
relationship, we raise the index $\rho$ in Eq.~(\ref{eq grav 2}) by
contracting again with Eq.~(\ref{eq grav 1}); thus it results that
\begin{equation}
 \mathcal{G}~\frac{(\breve{F}^\mu_{\ \rho }-\beta ^{2}\;\breve{P}\ ^{\ast }\breve{F}_{~\rho }^{\mu })~(\mathcal{\bar{G}}^{\rho \nu }+\beta ^{2}~^{\ast }\breve{F}_{~\lambda }^{\rho }~^{\ast }\breve{F}
^{\lambda \nu })}{1+2\beta ^{2}~\breve{S}-\beta ^{4}~\breve{P}^{2}}
= \lambda^2 g\ F^{\mu\rho} \ ,
\end{equation}
which, via Eq.~(\ref{tripleF}) is converted into
\begin{equation}\label{eq grav 3}
 \mathcal{G}~\frac{(1+\beta^4\breve{P}^2)~\breve{F}^{\mu\nu }-2\beta ^{2}\;\breve{P}~(1+\beta^2\breve{S})\ ^{\ast }\breve{F}^{\mu\nu } }{1+2\beta
^{2}~\breve{S}-\beta ^{4}~\breve{P}^{2}} = \lambda^2 g\ F^{\mu\nu} \
.
\end{equation}
We can also find the relationship between dual matrices. For this,
we contract the former equation with the Levi-Civita symbol
$\epsilon_{\mu\nu\lambda\rho}$, which must be associated with
factors $\sqrt{-\mathcal G}$ and $\sqrt{-g}$ already present in the
l.h.s. and the r.h.s., respectively:
\begin{equation}\label{eq grav 4}
\sqrt{\mathcal{G}}~~\frac{(1+\beta^4\breve{P}^2)~^{\ast
}\breve{F}_{\lambda\rho}+2\beta
^{2}\;\breve{P}~(1+\beta^2\breve{S})\ \breve{F}_{\lambda\rho}
}{1+2\beta ^{2}~\breve{S}-\beta ^{4}~\breve{P}^{2}} = \lambda^2
\sqrt{g}~~ ^{\ast }F_{\lambda\rho} \ .
\end{equation}
We are ready to get the relationship between the field invariants.
By contracting Eq.~(\ref{eq grav 3}) with Eq.~(\ref{eq grav 4}),
and then taking the trace, the result is
\begin{equation}
(- \mathcal{G})^{3/2}~\frac{(1+\beta^4\breve{P}^2)^2-4\beta
^{4}\;\breve{P}^2~(1+\beta^2\breve{S})^2+4\beta^2\  \breve{S}\
(1+\beta^4\breve{P}^2)(1+\beta^2\breve{S}) }{(1+2\beta
^{2}~\breve{S}-\beta ^{4}~\breve{P}^{2})^2} \ \breve{P}= \lambda^4
(-g)^{3/2}\ P \ ,
\end{equation}
which simplifies to yield
\begin{equation}
(- \mathcal{G})^{3/2}~\breve{P}= \lambda^4 (-g)^{3/2}\ P \ ,
\end{equation}
and together with Eq.~(\ref{relation_PP}) leads to the dynamical
relationships between the pseudoscalars and the volumes,
\begin{equation}\label{PPGg}
   \lambda^2\ \breve{P}=P\ ,\ \ \ \ \ \ \ \ \sqrt{-\mathcal G}=\lambda ^2 \sqrt{-g}\ ;
\end{equation}
the second relationship implies that the ``metric'' $\mathcal G$ is inversible if the metric $g$ is. In this way, the general relation (\ref{dualdual}) is dynamically
converted into
\begin{equation}\label{dualdual 2}
  \lambda^2\  ^{\ast}\breve{F}^{\mu\lambda }=\ ^{\ast}{F}^{\mu\lambda }~.
\end{equation}
On the other hand, by contracting Eq.~(\ref{eq grav 3}) with
$F_{\mu\nu}$, then taking the trace and using
$\mathcal{G}=\lambda^4\ g$, one obtains the expression of $S$ in
terms of $\breve S$ and $\breve P$:
\begin{equation}\label{S 2}
\lambda^2~\frac{\breve{S}\ (1-\beta ^{4}\;\breve{P}^2)-2\ \beta^2\
\breve{P}^2}{1+2\beta ^{2}~\breve{S}-\beta ^{4}~\breve{P}^{2}} =S \
.
\end{equation}
Now, let us focus on a calculation concerning the structure of
BI field equation (\ref{cursive F bornInfeld}).
Eqs.~(\ref{dualdual 2}), (\ref{eq grav 3}) and (\ref{PPGg}) can be
combined to yield
\begin{eqnarray}\label{eq em 3}
    {F}^{\mu\nu }+\beta^2  \lambda^{-2}P\; ^{\ast}{F}^{\mu\nu }&=&\lambda^2 \frac{(1+\beta^4\breve{P}^2)\breve{F}^{\mu\nu }-2\beta ^{2}\breve{P}(1+\beta^2\breve{S}) ^{\ast }\breve{F}^{\mu\nu } }{1+2\beta
^{2}~\breve{S}-\beta ^{4}~\breve{P}^{2}}+\lambda^2\
\beta^2\breve{P}\ ^{\ast}\breve{F}^{\mu\nu }\notag\\ \notag\\
&=&\lambda^2 \frac{(1+\beta^4\breve{P}^2)~(\breve{F}^{\mu\nu
}-\beta^2\breve{P}\ ^{\ast}\breve{F}^{\mu\nu
})}{1+2\beta^{2}~\breve{S}-\beta ^{4}~\breve{P}^{2}}\ .
\end{eqnarray}
Notice that Eq.~(\ref{S 2}) implies
\begin{equation}\label{conversion}
    1-2(\beta/\lambda)^{2}\; S-(\beta/\lambda)^{4}\; P^2=1-2\beta^2\ \left(\frac{\breve{S}\ (1-\beta ^{4}\;\breve{P}^2)-2\ \beta^2\ \breve{P}^2}{1+2\beta
^{2}~\breve{S}-\beta ^{4}~\breve{P}^{2}} \right)-\beta^4
\breve{P}^2=\frac{(1+b^{-4}\breve{P}^2)^2}{1+2\beta
^{2}~\breve{S}-\beta ^{4}~\breve{P}^{2}}\ ,
\end{equation}
which allows Eq.~(\ref{eq em 3}) to be written as
\begin{equation}\label{eq em 2}
  \lambda^{-2}\  \frac{{F}^{\mu\nu }+(\beta/  \lambda)^{2}\; P\ ^{\ast}{F}^{\mu\nu }}{\sqrt{1-2(\beta/\lambda)^{2}\; S-(\beta/\lambda)^{4}\; P^2}} =\frac{\breve{F}^{\mu\nu }-\beta^2\breve{P}\ ^{\ast}\breve{F}^{\mu\nu }}{\sqrt{1+2\beta^{2}~\breve{S}-\beta ^{4}~\breve{P}^{2}}}\ .
\end{equation}
This equation can be multiplied by the volume $\sqrt{-\mathcal G}$ (so
$\sqrt{-g}$ will appear in the l.h.s., according to
Eq.~(\ref{PPGg})) to get on the right side the density tensor that
enters the electrodynamic equation (\ref{eq em 0}).

Equation (\ref{eq em 2}) is a direct consequence of applying the
dynamical equation (\ref{eq grav 1}). It has a profound meaning; the
electrodynamics that emerges from Lagrangian (\ref{eq accion}) can
be understood both as the standard BI electrodynamics in
the geometric background of ``metric''
$\mathcal{G}_{\mu\nu}=g_{\mu\nu}+\epsilon R_{\mu\nu} $ (where the
electromagnetic field seems to be coupled to the curvature), or as
an \textit{anomalous} BI electrodynamics in the geometric
background of metric $g_{\mu\nu}$,
\begin{equation} \label{eq em final}
\boxed{
\partial_\mu \left( \sqrt{-{g}} \   \frac{{F}^{\mu\nu }+(\beta/  \lambda)^{2}\; P\ ^{\ast}{F}^{\mu\nu }}{\sqrt{1-2(\beta/\lambda)^{2}\; S-(\beta/\lambda)^{4}\; P^2}} \right) = 0\ .}
\end{equation}
We distinguish these two possibilities by calling them
\textit{picture} $\mathcal G$ and \textit{picture} $g$ respectively.
We say that the picture $g$ is anomalous because $S$ appears with
the ``wrong'' sign in the l.h.s. of Eq.~(\ref{eq em 2}) . Therefore, for
passing from the (standard) r.h.s. to the (anomalous) l.h.s. in
Eq.~(\ref{eq em 2}) not only the metric must be changed, but
$\beta^2$ must be replaced with $-(\beta/\lambda)^{2}$ (or,
alternatively, $\beta^2\rightarrow -\beta^2 $,
$F_{\mu\nu}\rightarrow \lambda^{-1}F_{\mu\nu}$). However, the
picture $g$ is not intrinsically anomalous, since we could have
started from a Lagrangian built with an imaginary value for $\beta$;
in this way the anomalous character would be transferred to the
picture $\mathcal G$, while the l.h.s of  Eq.~(\ref{eq em 2})  would
look as the standard BI electrodynamics for the field
$\lambda^{-1}F_{\mu\nu}$.  Imaginary values for $\beta$ are not
forbidden in this theory, since only even powers of $\beta$ appear
in the dynamical equations (the Lagrangian (\ref{eq accion}) is made
of even powers of $\beta$ according to Eq.~(\ref{detq}) ).

Let us now focus on the gravitodynamic equation (\ref{eq grav 0}),
where the Ricci tensor of the geometry $g_{\mu\nu}$ is sourced by an
energy-momentum built in the picture $\mathcal G$, which implies
that the curvature is on both sides of the equation. So we wonder
whether it is possible to give this equation a form in picture $g$,
where the curvature is sourced by an energy momentum that does not
depend on the curvature. First, note that $\det(
\delta^{\mu}_{\nu}+\epsilon\, R^\mu_{\ \nu})=\mathcal{G}/g$ is
dynamically equal to $\lambda^4$ (see Eq.~(\ref{PPGg})), so it will
be easy to rewrite Eq.~(\ref{eq grav 0})  in terms of the Einstein
tensor. In particular, the curvature scalar can be computed by
taking the trace to the equation:
\begin{equation}
  \lambda^{-1}\, (4+\epsilon\, R)=\frac{4+\beta ^{2}~4 \breve{S}}{\sqrt{1+2\beta
^{2}~\breve{S}-\beta ^{4}~\breve{P}^{2}}}\ .
\end{equation}
Therefore, by using (\ref{relacion_S}) and other properties already
used, one gets
\begin{equation}\label{grav}
   \lambda^{-1} \epsilon\; (G^\mu_\nu-\Lambda\; \delta^\mu_\nu)= \lambda^{-1}\epsilon\; (R^\mu_\nu-\frac{1}{2}\; R\; \delta^\mu_\nu)-( \lambda^{-1}-1)\;\delta^\mu_\nu=\delta^\mu_\nu-\frac{\delta^\mu_\nu-\beta^2\ \breve{F}^{\mu\lambda}\; F_{\lambda\nu}}{\sqrt{1+2\beta
^{2}~\breve{S}-\beta ^{4}~\breve{P}^{2}}}\ .
\end{equation}
The r.h.s.~is necessarily a conserved tensor. However, it is not the
electromagnetic energy-momentum tensor in picture $\mathcal{G}$ (the l.h.s.either is the Einstein tensor in that picture).
Therefore, we will try to give the r.h.s.~the form of the
energy-momentum electromagnetic tensor in the picture $g$. The
energy-momentum (\ref{T munu BI}) of standard BI
electrodynamics can be rewritten, among other equivalent forms, as
\begin{equation}
    4\pi\beta^2\ T^\mu_\nu=- \delta^\mu_\nu+\frac{\delta^\mu_\nu+\beta^2~^{\ast }{F}^{\mu\lambda}\; ^{\ast}{F}_{\lambda\nu}}{\sqrt{1+2\beta
^{2}~{S}-\beta ^{4}~{P}^{2}}}\ .
\end{equation}
For the anomalous case, it then results in
\begin{equation}\label{Tanom}
   - 4\pi(\beta/\lambda)^2\ {T^{anom}}^\mu_{\;\nu}=- \delta^\mu_\nu+\frac{\delta^\mu_\nu-(\beta/\lambda)^2~^{\ast }{F}^{\mu\lambda}\; ^{\ast}{F}_{\lambda\nu}}{\sqrt{1-2(\beta/\lambda)
^{2}~{S}-(\beta/\lambda) ^{4}~{P}^{2}}}\ ,
\end{equation}
which can be translated into the variables of picture $\mathcal{G}$
by means of (\ref{eq grav 4}) , (\ref{dualdual 2}) and
(\ref{relacion_P}),
\begin{align}
    \delta_\mu^\nu-(\beta/\lambda)^{2}\ ^{\ast}{F}^{\mu\lambda}\; ^{\ast}{F}_{\lambda\nu}&=  \delta_\mu^\nu-\beta^{2}\ \frac{(1+\beta^4\breve{P}^2)~^{\ast }\breve{F}^{\mu\lambda}\; ^{\ast}\breve{F}_{\lambda\nu}-2\beta ^{2}\;\breve{P}^2~(1+\beta^2\breve{S})\ \delta^\mu_\nu }{1+2\beta
^{2}~\breve{S}-\beta ^{4}~\breve{P}^{2}}\notag\\ \notag\\ &=
\delta^\mu_\nu\ \frac{1+2\beta ^{2}\breve{S}-\beta
^{4}\breve{P}^{2}+2\beta^4 \breve{P}^2(1+\beta^2\breve{S})}{1+2\beta
^{2}~\breve{S}-\beta
^{4}~\breve{P}^{2}}-\frac{(1+\beta^4\breve{P}^2)~\beta^{2}~^{\ast
}\breve{F}^{\mu\lambda}\; ^{\ast}\breve{F}_{\lambda\nu} }{1+2\beta
^{2}~\breve{S}-\beta ^{4}~\breve{P}^{2}}\notag\\ \notag\\ &=
\frac{1+\beta ^{4}\breve{P}^2}{1+2\beta ^{2}~\breve{S}-\beta
^{4}~\breve{P}^{2}}\ \left((1+2\beta ^{2}\breve{S})\
\delta^\mu_\nu-\beta^{2}~^{\ast }\breve{F}^{\mu\lambda}\;
^{\ast}\breve{F}_{\lambda\nu} \right) \notag\\ \notag\\ &=
\frac{1+\beta ^{4}\breve{P}^2}{1+2\beta ^{2}~\breve{S}-\beta
^{4}~\breve{P}^{2}}\
\left(\delta^\mu_\nu-\beta^{2}~\breve{F}^{\mu\lambda}\;
{F}_{\lambda\nu} \right) \ .
\end{align}
Therefore, by applying Eq.~(\ref{conversion}) it results that
\begin{equation}
    \frac{\delta_\mu^\nu-(\beta/\lambda)^{2}\ ^{\ast}{F}^{\mu\lambda}\; ^{\ast}{F}_{\lambda\nu}}{\sqrt{1-2(\beta/\lambda)
^{2}~{S}-(\beta/\lambda)
^{4}~{P}^{2}}}=\frac{\delta^\mu_\nu-\beta^{2}~\breve{F}^{\mu\lambda}\;
{F}_{\lambda\nu} }{\sqrt{1+2\beta ^{2}~\breve{S}-\beta
^{4}~\breve{P}^{2}}}\ .
\end{equation}
Thus, the source in Eq.~(\ref{grav}) is the anomalous
energy-momentum tensor in picture $g$:
\begin{equation} \label{eq gravity final}
\boxed{
   G^\mu_\nu-\Lambda\; \delta^\mu_\nu=\frac{4\pi\beta^2}{\epsilon\lambda}\ {T^{anom}}^\mu_{\;\nu}\ .}
\end{equation}
As already stated, BI electrogravity depends only on the
even powers of $\beta$. In case an imaginary value for $\beta$ is
chosen --to avoid the anomalous behavior in the picture $g$-- the
sign of $\epsilon$ must be consistently changed to preserve the
positive sign of Newton's constant in the above equation (\textit{cf.} \cite{vollick06}). In such a
case, the cosmological constant in the former equation must be
understood as $(1-\lambda)/(-\epsilon)$.

\section{Dyonic Black Holes}
\label{cap:dionic_bh} Equations~(\ref{eq em final}) and (\ref{eq
gravity final}) make up an Einstein--Born--Infeld system that is
equivalent to the system formed by Eqs.~(\ref{eq em 0}) and (\ref{eq
grav 0}).  Its spherically symmetric solutions have been widely
studied, and we refer the reader to  \cite{review_beltran} (and
references therein) for a comprehensive review on this topic. This
geometry has been studied in standard Einstein--Born--Infeld contexts \cite{Salazar, Breton} and also in the anomalous determinantal context
\cite{afonso_et_al}.

We are interested in the capabilities of this type of theory for
smoothing singularities; as an example we solve the field equations
for the solution corresponding to a dyonic BH, where both
the electric and magnetic monopoles are present.
\subsection{Electrodynamics}
To work with Eq.~\eqref{eq em final}, we propose for the
electromagnetic field the closed 2-form
\begin{equation} \label{eq F_0 ionic}
\textbf{F} = - e(r) \ dt\wedge dr - g(\theta) \ d\theta \wedge
d\phi\ ,
\end{equation}
in a spherically symmetric geometric background given by
\begin{equation} \label{eq metric0 dionico}
   g_{\mu \nu}=  \text{diag}\left(
f(r),  -f^{-1}(r) ,-r{^2}, -r{^2} \sin{^2}\theta\right)\ .
\end{equation}
$f(r)$ is a function to be determined by the geometrodynamics, but
it decouples from the electrodynamical problem. In fact, $f(r)$ does
not enter into the volume $\sqrt{-g}$ nor into the operation of
raising indices to  $F_{\mu \nu}$ in Eq. ~(\ref{eq F_0 ionic}). The
field configuration (\ref{eq F_0 ionic}) describes an electrostatic
field $E_r=e(r)$ and a magnetostatic field
$B_r=r^{-2}g(\theta)\sin\theta $, both in the radial direction.  The
spherical symmetry imposes that $g(\theta ) = p \sin \theta$; thus
the magnetic field is Coulombian,
\begin{equation} \label{eq campo mag}
    B_r  = \frac{p}{r^2}\ .
\end{equation}

By replacing the field invariants $ S = (p^2 r^{-4}-e(r)^2)/2 $ and
$P = -r^{-2} p\ e(r)$, Eq. \eqref{eq em final} is integrated to
yield
\begin{equation} \label{eq campo electrico dionico}
     e(r) = \frac{q}{\sqrt{r^4-(\beta /\lambda)^2 (q^2+p^2)}}\, .
\end{equation}
If $\beta$ is real in the Lagrangian (i.e., picture $g$ is anomalous), then \( e(r) \) is real only for $r>r_o\equiv
\left[\beta^2\lambda^{-2}(q^2 + p^2)\right]^{1/4}$.  The magnetic field (\ref{eq
campo mag}) is regular in the domain $r \in(r_o,\infty)$, but $e(r)$
diverges at $r=r_o$,  Instead, if $\beta$ is imaginary in the Lagrangian (i.e., picture $g$ is standard), then the solution is defined in the domain $r\in (0,\infty)$; the electric field is
regular everywhere (it is the standard BI field of a
point-like charge), but the magnetic field diverges at $r=0$.

The energy-momentum tensor (\ref{Tanom}) is
\begin{equation}\label{Tanom dionico}
   {T^{anom}}^\mu_{\;\nu}=\frac{\lambda^2}{4\pi\ \beta^2}\ \text{diag}\left(1-\sqrt{1-\frac{r_o^4}{r^4}} \ ,\ 1-\sqrt{1-\frac{r_o^4}{r^4}}\ ,\ 1-\frac{1}{\sqrt{1-\frac{r_o^4}{r^4}}}\ ,\ 1-\frac{1}{\sqrt{1-\frac{r_o^4}{r^4}}}\right)\ ,
\end{equation}
whose trace determines the behavior of the curvature scalar $R$ in
Eq.~(\ref{eq gravity final}):
\begin{equation}\label{curvatura escalar dionico}
 R=-4\ \frac{1-\lambda}{\epsilon}+\frac{2\lambda}{\epsilon}\  \frac{\left(1-\sqrt{1-\frac{r_o^4}{r^4}}\right)^2}{\sqrt{1-\frac{r_o^4}{r^4}}}\ .
\end{equation}
Note that even though the electric and magnetic monopoles $q$ and
$p$ enter the field invariants $S$ and $P$ in an unequal way, they
contribute to the energy and pressure on an equal footing through
the combination $\sqrt{q^2+p^2}$.  This feature will be inherited by
the geometry we are going to examine.

 \subsection{Gravitodynamics}
The Ricci tensor  \(R^\mu_{\ \nu}\) for the metric (\ref{eq metric0
dionico}) is
\begin{equation} \label{eq Ricci dionico} R^\mu_{\ \nu}=
 \text{diag}\left(
 \frac{2 f^{\prime } + r f^{\prime \prime}}{2r },\frac{2 f^{\prime } + r f^{\prime \prime}}{2r }, - \frac{1-f(r)-rf^{\prime}}{r^2} , - \frac{1-f(r)-rf^{\prime}}{r^2}\right),
\end{equation}
To solve the gravitodynamic equation (\ref{eq gravity final}) with the source (\ref{Tanom dionico}), $f$ must be a solution to the first order equation
\begin{equation} \label{eq gravity dionico}
    -\left(\frac{r}{r_o}\right)^2 + \lambda \ \sqrt{\left(\frac{r}{r_o}\right)^4-1} - \frac{\epsilon}{r_o^2} \left[-1 + f(r) + r f^{\prime} (r)\right] = 0\ .
\end{equation}
It will be convenient to switch to nondimensional coordinate and parameter:
\begin{equation}\label{alpha2}
    \rho\equiv\frac{r}{r_o}=\frac{\sqrt{\lambda\beta^{-1}}\ r}{(q^2+p^2)^{1/4}}\ ,\ \ \ \ \  \alpha \equiv \frac{\epsilon}{r_o^2}=\frac{\beta}{2G\sqrt{q^2+p^2}}\ ,
\end{equation}where the relationship $G=\beta^2/(2\lambda\epsilon)$ emerged from the Eq.~(\ref{eq gravity final}). Thus, the solution to Eq.~(\ref{eq gravity dionico}) reads
\begin{equation} \label{eq solucion dionico}
    f(\rho) = 1 - \frac{C}{\rho} - \frac{\rho^2}{3\ \alpha}  + \frac{\lambda}{3\alpha}\ \sqrt{\rho^4 -1}
+ \frac{2\lambda}{3\alpha\ \rho^2 }  \,\ {_2 F_1}\left(\frac{1}{4},\frac{1}{2};\frac{5}{4};\rho^{-4}\right)\ .
\end{equation}
Here, ${_2 F_1}(\frac{1}{4},\frac{1}{2};\frac{5}{4};\frac{1}{x})$ is the ordinary hypergeometric function, which is real and finite for $x\in(-\infty,0]\cup [1,\infty)$. Therefore, the function \eqref{eq solucion dionico} is defined in the domain $\rho\ge1$.\footnote{For further checks, the hypergeometric function satisfies ${_2 F_1}(\frac{1}{4},\frac{1}{2};\frac{5}{4};\rho^{-4})=\rho\,  [ 2 i \ \text{K}(-1) +\text{K}(2) - i \ \text{F}(\arcsin\rho|-1)]$, where $\text{F}(\varphi|m)$ is the elliptic integral of the first
kind, and $\text{K}(m)$ is the complete elliptic integral of the
first kind. In addition, ${_2 F_1}(\frac{1}{4},\frac{1}{2};\frac{5}{4};-\rho^{-4})=\rho\,\sqrt{i} \left[ 2 \ \text{K}(-1) - i \ \text{K}(2) + \ \text{F}(i\ \sinh^{-1} (\sqrt{i}\ \rho)|-1) \right]$.} $C$ is an integration constant proportional to the BH mass,
\begin{equation}\label{eq:mass}
    M=\frac{r_o}{2G}\ C=\frac{\sqrt{\epsilon}}{2G\sqrt{\alpha}}\ C=\frac{\beta}{(2G)^{3/2}\sqrt{\alpha}}\ C\ .
\end{equation}

Although $f(\rho)$ and $f'(\rho)$ are finite at $\rho=1$, the geometry will be singular there because $f''(\rho=1)$ diverges. Furthermore, the behavior of $f(r)$ at infinity is
\begin{equation}\label{fdio}
    f(r)=1-\frac{2MG}{r}-\frac{\Lambda}{3}\ r^2+\frac{G\, (q^2 + p^2)}{r^2}+\frac{\beta^2G(q^2 + p^2)^2}{20\lambda^2\, r^6} +\mathcal{O}(r^{-10})\ .
\end{equation}
On the other hand, if
$(\beta,\epsilon)$ in the Lagrangian are replaced with $(i\beta, -\epsilon)$,\footnote{$\epsilon$ changes sign to  preserve the Newton constant
in Eq.~(\ref{eq gravity final}). The cosmological constant $\Lambda$
in Eq.~(\ref{eq gravity final}) is now equal to
$(1-\lambda)/(-\epsilon)$.} while keeping the positive values
of the (new) parameters $\beta$, $\epsilon$, then the electric field
$e(r)$ becomes regular at the origin (as already shown), and the equation for $f$ becomes
\begin{equation} \label{eq gravity dionico imaginario}
    -\left(\frac{r}{r_o}\right)^2 + \lambda \ \sqrt{\left(\frac{r}{r_o}\right)^4+1} + \frac{\epsilon}{r_o^2} \left[-1 + f(r) + r f^{\prime} (r)\right] = 0\ ,
\end{equation}
which admits the solution
\begin{equation} \label{eq solucion dionico imaginario}
    f(\rho) = 1 - \frac{C}{\rho} + \frac{\rho^2}{3\ \alpha}  - \frac{\lambda}{3\alpha}\ \sqrt{\rho^4 +1}
+ \frac{2\lambda}{3\alpha\ \rho^2 }  \,\ {_2 F_1}\left(\frac{1}{4},\frac{1}{2};\frac{5}{4};-\rho^{-4}\right)\ .
\end{equation}
This is a real function throughout the domain $\rho>0$. The behavior of $f(r)$ at infinity is
\begin{equation}\label{fdioimaginario}
    f(r)=1-\frac{2MG}{r}-\frac{\Lambda}{3}\ r^2+\frac{G\, (q^2 + p^2)}{r^2}-\frac{\beta^2G(q^2 + p^2)^2}{20\lambda^2\, r^6} +\mathcal{O}(r^{-10})\ .
\end{equation}
Comparing the metrics (\ref{fdio}) and (\ref{fdioimaginario}), we see that the choice of an imaginary parameter $\beta$ in the Lagrangian changes the metric to just the order $r^{-6}$, which is also the order in which these metrics depart from the Reissner-Nordstr\"{o}m metric,
\begin{equation}
    f_{RN}(r)=1-\frac{2MG}{r}-\frac{\Lambda}{3}\ r^2+\frac{G\, Q^2}{r^2}\ .
\end{equation}

\subsection{Horizon structure and extremal black holes}

In what follows, we restrict our analysis to asymptotically flat
solutions by fixing $\lambda=1$, so that the cosmological constant
vanishes.

The properties of the spacetime studied here are determined entirely by the behavior of the metric function $f(\rho)$, whose profile depends on the integration constant $C$ and the dimensionless parameter $\alpha$. Depending on the values of these parameters, $f(\rho)$ may exhibit different concavity properties, develop a minimum or not, and lead to one, two, or no horizons. We begin by discussing these qualitative behaviors before deriving
the analytical conditions for extremality.

\begin{figure}
\hspace{-0.2cm}
\begin{subfigure}{0.49\linewidth}
    \centering
    \includegraphics[width=\linewidth]{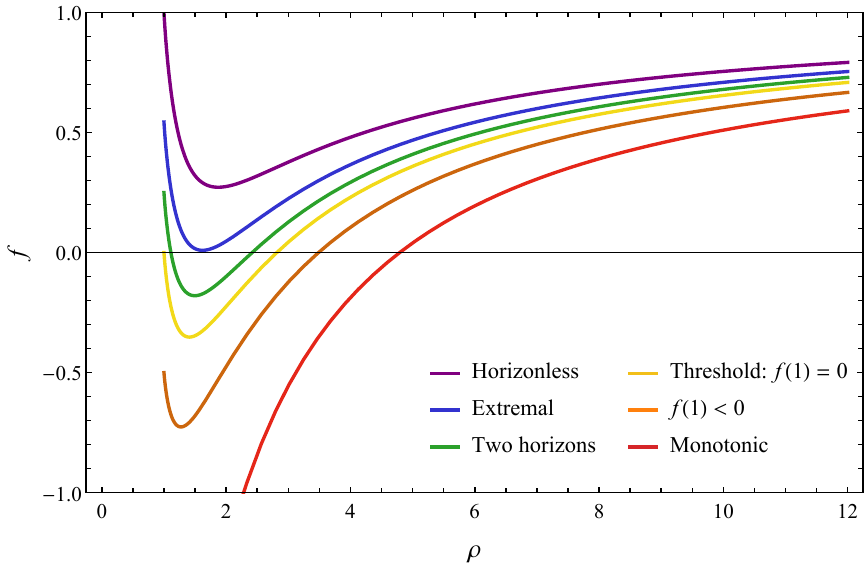}
    \caption{}
    \label{fig:c0}
\end{subfigure}
\hspace{-0.1cm}
\begin{subfigure}{0.48\linewidth}
    \centering
    \includegraphics[width=\linewidth]{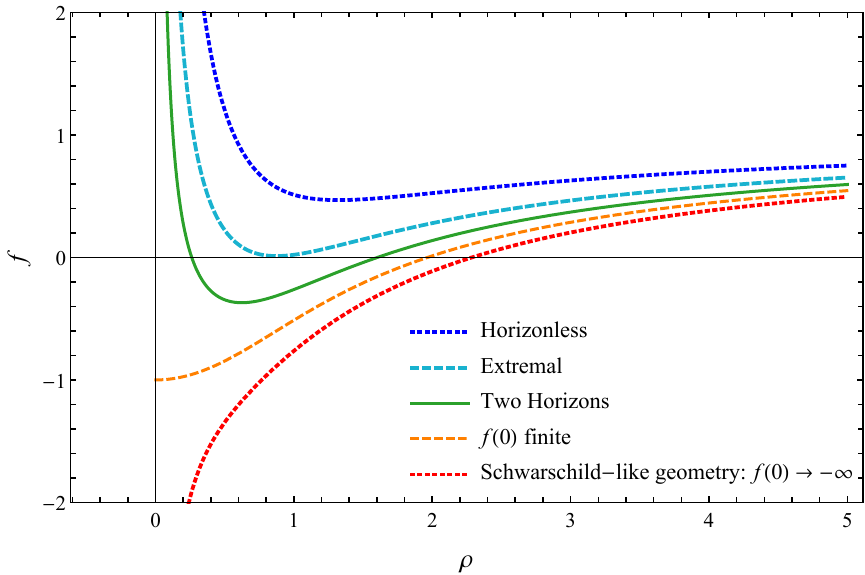}
    \caption{}
    \label{fig:f_rn_schw}
\end{subfigure}
\caption{\textbf{Left:} Function $f(\rho)$ in Eq.~\eqref{eq solucion dionico}, with $\alpha=0.2$. The curves illustrate all
possible configurations: a horizonless geometry (purple), an extremal black hole
(blue), a black hole with two horizons (green), the threshold case satisfying
$f(1)=0$ (yellow), a single-horizon black hole with $f(1)<0$ (orange), and the
monotonic single-horizon case (red). \textbf{Right:}  Function $f(\rho)$ in Eq.~\eqref{eq solucion dionico imaginario}, with $\alpha=0.5$. The orange dashed curve,
obtained by setting $C=C_{\rm crit}$ (Eq.~\eqref{eq:c_crit}), marks the
threshold separating ``Schwarzschild-like'' geometries
(red dotted curve), for which $f(\rho\rightarrow0)\rightarrow-\infty$, from ``Reissner--Nordstr\"om-like'' geometries,
for which $f(\rho \rightarrow0)\rightarrow+\infty$. In the latter case, the metric
may describe a black hole with two horizons (green curve), an extremal
black hole (light-blue dashed curve), or a horizonless geometry (blue
dotted curve), depending on the value of $C$.}
\end{figure}

To locate the horizons of the metric one searches for the roots of
$f(\rho)=g_{00}$. In Reissner--Nordström
geometry, $f(\rho)$ has two roots or none; the special case of degenerate roots is the \textit{extremal} black hole. If the roots are different, the BH displays a (Cauchy) inner horizon and an (event) outer horizon. Cauchy horizons are unstable under generic perturbations due to the infinite blueshift experienced by ingoing radiation, a mechanism that may lead to mass inflation and the formation of a null singularity \cite{PoissonIsrael90}. One of the main reasons for combining general relativity and nonlinear electrodynamics is the search for (monotonic) solutions $f(\rho)$ with a single root, to avoid such a type of instabilities \cite{hale}. These Schwarzschild-like BHs are known to exist in standard BI electrodynamics \cite{oliveira, breton01} (see Figure~\ref{fig:f_rn_schw}). For the anomalous BI electrodynamics, Figure~\ref{fig:c0} summarizes the different qualitative behaviors of the metric function $f(\rho)$ in Eq.~\eqref{eq solucion dionico}. A spacetime with a single horizon may arise in two different ways, since the monotonic behavior is not the unique way for $f(\rho)$ to intersect the horizontal axis only once. In fact,  $f(\rho)$ may develop a local minimum while satisfying $f(1)<0$; in this case the inner root lies outside the physical domain $\rho\geq1$, leaving only the outer horizon. This other possibility will impact on the thermodynamical behavior, which will be studied in the following subsection. If $f(1)>0$, then the geometry will have two or no horizons, the extremal case being the separatrix between both behaviors. $f(1)\lessgtr 0$ implies a condition on the relationship between $C$ and $\alpha$ in Eq.~\eqref{eq solucion dionico}, 
\begin{equation} \label{eq:c_0}
    f(1)\lessgtr 0\ \ \ \ \ \iff\ \ \ \ \ C\gtrless 1-\frac{1}{3\alpha}\, 
 \left(1-2\sqrt{\pi}\ \frac{\Gamma\left(\frac{5}{4}\right)}{\Gamma\left(\frac{3}{4}\right)}\right) \equiv C_0\ .
\end{equation}
The condition to have a monotonic behavior is
\begin{equation}
    f'(\rho)>0\, ,\ \forall \rho\ \ \ \ \ \iff\ \ \ \ \ C>\frac{2}{3\alpha}\, \left(1+\sqrt{\pi}\, \frac{\Gamma\left(\frac{5}{4}\right)}{\Gamma\left(\frac{3}{4}\right)}\right)\equiv C_1\ .
\end{equation}
In any case, as long as there is at least one horizon, the electric field divergence at $\rho=1$ remains hidden behind it. More interestingly, in configurations with two horizons, if the Cauchy horizon undergoes mass inflation, as expected by analogy with the Reissner--Nordström solution, the resulting null singularity could prevent a macroscopic observer from ever reaching the electrodynamic breakdown at $\rho=1$. A detailed analysis of this issue, however, lies beyond the scope of the present work.
\bigskip

If the determinantal Lagrangian is built with an imaginary
value of $\beta$, then the geometry will be characterized by
$f(\rho)$ in Eq.~(\ref{eq solucion dionico imaginario}). Although this function reproduces the Reissner-Nordstr\"{o}m geometry at infinity (see Eq.~\eqref{fdioimaginario}), its behavior at $\rho=0$ is strongly dependent on the choice of $C$. This is because the term containing the hypergeometric function competes with the mass term in the limit $\rho\rightarrow 0$. In fact, the sign of the divergence of $f(\rho)$ at $\rho=0$ satisfies
\begin{equation}\label{eq:c_crit}    
f(\rho)\xrightarrow[\rho\to 0]{} \pm\infty\ \ \ \ \iff\ \ \ \ C\lessgtr  \frac{2\, \Gamma(\frac{1}{4})\Gamma(\frac{5}{4})}{3\sqrt{\pi}\,\alpha}\equiv C_{\text{crit}}\ .
\end{equation}
These two ways of choosing $C$ separate the ``Schwarzschild-like'' behavior ($f(\rho\rightarrow 0)\rightarrow -\infty$) from the ``Reissner-Nordstr\"{o}m-like'' behavior ($f(\rho\rightarrow 0)\rightarrow +\infty$). If $C=C_{\text{crit}}$ then $f(\rho=0)$ is finite. The different behaviors are shown in Figure \ref{fig:f_rn_schw}. In both cases, the magnetic field divergence at $\rho = 0$ is shielded by a horizon. 
\bigskip

\bigskip

\textbf{Extremal BHs.} In Reissner-Nordstr\"{o}m geometry, the roots of $f$ coincide when $GM^2=Q^2$. In this case, the black hole is extremal and the
Cauchy horizon merges with the event horizon. In the geometry studied here, the
roots of $f$ cannot be obtained analytically. However, the extremal case can still be identified by requiring that
both $f$ and $f'$ vanish at the same point $\rho_h$. This double condition results in two
equations to find both $\rho_h$ and the relation $C=C(\alpha)$. Thus, in Reissner-Nordstr\"{o}m one obtains
\begin{equation}
    1-\frac{2MG}{r_h}+\frac{G Q^2}{r_h^2}=0\ ,\ \  \ \ \frac{2MG}{r_h^2}-2\frac{G Q^2}{r_h^3}=0\ \ \  \ \Rightarrow \ \  \ \sqrt{G}M=|Q| \ , \  \ \ \ r_h=G M\ .
\end{equation}
At the level of nondimensional variables, this result corresponds to $C=\sqrt{2/\alpha}$,
$\rho_h=C/2$  (or
$\rho_h=1/\sqrt{2\alpha}$).

Applying this criterion to the function (\ref{eq solucion dionico}) ---the solution with real $\beta$ in the Lagrangian--- we find
\begin{equation} \label{eq:extremal_horizon}
    \rho_h=\frac{\sqrt{1+\alpha^2}}{\sqrt{2\alpha}}\ , \ \ \ \ \ \ \ \ \ \ C_{\text{extr}}=\sqrt{\frac{2}{\alpha}}\ \frac{\sqrt{1+\alpha^2}}{3}
\left[1+\frac{2}{1+\alpha^2}  \,\ {_2 F_1}\left(\frac{1}{4},\frac{1}{2};\frac{5}{4};\left(\frac{2\alpha}{1+\alpha^2}\right)^2\right)\right]\ .
\end{equation}
The function $\rho_h(\alpha)$ has two branches, which are related by the equality $\rho_h(\alpha)=\rho_h(\alpha^{-1})$. However, only the branch belonging to the interval $\alpha\in(0,1]$ satisfies both conditions, $f(\rho_h)=0=f'(\rho_h)$, leading to the extremal BH.

As Eq.~\eqref{eq:extremal_horizon} shows, the nondimensional extremal horizon radius has a lower bound of $\rho_h \geq 1$, which is reached at $\alpha = 1$. This ensures that the electric field divergence at $\rho = 1$ is behind the horizon (or coincident with it for $\alpha=1$), thus preventing a naked electromagnetic singularity.

Concerning the extremal BH for the geometries described by the function $f(\rho)$ in Eq.~(\ref{eq solucion dionico imaginario}), the conditions $f(\rho_h)=0=f'(\rho_h)$ are satisfied if
\begin{equation} \label{eq C dioimaginario}
   \rho_h=\frac{\sqrt{1-\alpha^2}}{\sqrt{2\alpha}}\ ,\ \ \ \ \ \ \ C_{\text{extr}} =\sqrt{\frac{2}{\alpha}}\ \frac{\sqrt{1-\alpha^2}}{3}
\left[1+\frac{2}{1-\alpha^2}  \,\ {_2 F_1}\left(\frac{1}{4},\frac{1}{2};\frac{5}{4};-\left(\frac{2\alpha}{1-\alpha^2}\right)^2\right)\right]\ ,
\end{equation}
where $\alpha$ must satisfy $\alpha \in (0,1]$. Both extremal BHs, \eqref{eq:extremal_horizon} and \eqref{eq C dioimaginario},  tend to the extremal Reissner-Nordstr\"{o}m BH in the Maxwellian limit \(\beta\rightarrow 0 \) (i.e., $\alpha\rightarrow 0$). In fact, it is ${_2 F_1}\left(\frac{1}{4},\frac{1}{2};\frac{5}{4};0\right)=1$; then
\begin{equation}
  \rho_h\xrightarrow[\beta\to 0]{}\frac{1}{\sqrt{2\alpha}} \ , \ \ \ \ \ \ \ \ \ \ C_{\text{extr}}\xrightarrow[\beta\to 0]{} \sqrt{
    \frac{2}{\alpha} }\ .
\end{equation}

Very interestingly, both the anomalous and the standard cases allow for an extremal BH whose horizon is located at the singularity ($\rho=1$ and $\rho=0$, respectively). This happens for $\alpha=1$; i.e., for the charge $Q\equiv \sqrt{q^2+p^2}=\beta/(2G)$ (the minimal charge the extremal BH can have, as seen in Eq.~\eqref{alpha2}). The respective masses are obtained by computing $C_{\text{extr}}(\alpha=1)$ for the anomalous case of Eq.~\eqref{eq:extremal_horizon} and the standard case of Eq.~\eqref{eq C dioimaginario}. In the first case it is $C_{\text{extr}}(\alpha=1)=C_0(\alpha=1)=C_1(\alpha=1)$; in the second case it is $C_{\text{extr}}(\alpha\to 1)= C_{\text{crit}}(\alpha=1)$. This means that in both cases the BH is at the transition between Schwarzschild and RN behaviors. By replacing these values for $C$ in Eq.~\eqref{eq:mass}, we obtain the respective BH masses:
\begin{equation}\label{Bornpart}
   M^{anom}_{\text{extr}}(\alpha=1)=\frac{\beta}{(2G)^{3/2}}\ \frac{2}{3}\left(1+\frac{\sqrt{\pi}\, \Gamma\left(\frac{5}{4}\right)}{\Gamma\left(\frac{3}{4}\right)}\right)\ ,  \ \ \ \ \ \ \ \ \ \ \ \ \ \ \ \ \ M^{std}_{\text{extr}}(\alpha=1)=\frac{\beta}{(2G)^{3/2}}\ \frac{2\, \Gamma(\frac{1}{4})\Gamma(\frac{5}{4})}{3\sqrt{\pi}}\ .
\end{equation}
These masses can be compared with the respective electromagnetic energies,
\begin{equation}\label{energy}
   U^{anom}=4\pi r_o^3\int_1^\infty T^{anom\ 0}_{\ \ \ \ \ \ \ \ 0}\ \rho^2 d\rho\ , \ \ \ \ \ \ \ \ \ \ \ \ \ \ \ \ \ U^{std}=4\pi r_o^3\int_0^\infty T^{std\ 0}_{\ \ \ \ \ 0}\ \rho^2 d\rho\ ,
\end{equation}
where $T^{anom\ 0}_{\ \ \ \ \ \ \ \ 0}$ is given in Eq.~\eqref{Tanom dionico}, and $T^{std\ 0}_{\ \ \ \ \ 0}=(4\pi\beta^2)^{-1}\left(-1+\sqrt{1+\rho^{-4}}\right)$ can be recovered by changing $\beta^2$ to $-\beta^2$. The results are
\begin{equation}
    M^{anom}_{\text{extr}}(\alpha=1)=U^{anom}+\frac{\beta}{(2G)^{3/2}}\ ,\ \ \ \ \ \ \ \ \ \ \ \ \ \ \ \ \ M^{std}_{\text{extr}}(\alpha=1)=U^{std}\ .
\end{equation}
In the standard case, the equality between mass and electromagnetic energy gives the BH the character of a ``Born particle'' \cite{russo}. In both cases, they are objects whose mass and charge are separately determined by the fundamental constants of the theory.

\subsection{Thermodynamics}
We now study the thermodynamic properties of the dyonic BH solutions. In particular, we derive the temperature, specific heat and entropy, and compare them with the corresponding Reissner--Nordström and Schwarzschild limits.

The Hawking temperature is $T=\kappa/(2\pi)$, where $\kappa$ is the horizon surface gravity. For metrics of the form (\ref{eq metric0 dionico}), the surface gravity can be computed as $\kappa=(1/2)  \, \partial f/\partial r$  (for example, see \cite{Ortin}). The derivative must be specialized at the outer horizon $r_+$, where $f(r_+)=0$. The extremal BH has $T=0$ because $\partial f/\partial r$ is zero at $r=r_h$.  

For the geometry (\ref{eq solucion dionico}) with $\lambda=1$, it results
\begin{equation}\label{T}
    T=\frac{1}{4\pi}\, \frac{\partial f}{\partial r}\Big\vert_{r=r_+}=\frac{1}{4\pi r_o}\, \frac{\partial f}{\partial \rho}\Big\vert_{\rho=\rho_+}=\frac{1}{4\pi r_o}\, \frac{\alpha-\rho_+^2+\sqrt{\rho_+^4-1}}{\alpha\,\rho_+}\ ,
\end{equation}
where $r_o=\beta\, (2G\, \alpha)^{-1/2}$ (see Eq.~\eqref{alpha2}). Although $\rho_+$ cannot be solved analytically, the temperature is manifestly positive for every non-extremal black hole, since $\partial f/\partial \rho>0$ at the outer horizon (see Figures \ref{fig:c0} and \ref{fig:f_rn_schw}). On the other hand, the mass $M=C r_o/(2G)$ can be written in terms of $\rho_+$ by solving $C$ from the equation $f(\rho_+)=0$; thus, the relation between $M$ and $\rho_+$ results to be
\begin{equation}\label{M}
    M=\frac{r_o}{6G\alpha}\, \left[-\rho_+^3+3\alpha\rho_+ +\rho_+\sqrt{\rho_+^4-1}+2\ {_2 F_1}\left(\frac{1}{4},\frac{1}{2};\frac{5}{4};\rho_+^{-4}\right)\right]\ .
\end{equation}
If $q^2+p^2$ goes to zero, then $\alpha$ tends to infinity (see Eq.~(\ref{alpha2})); therefore, the Schwarzschild relation $T=1/(8\pi GM)$ is obtained from Eqs. (\ref{T}) and (\ref{M}).

\begin{figure}
\hspace{-0.2cm}
\begin{subfigure}{0.49\linewidth}
    \centering
    \includegraphics[width=0.8\linewidth]{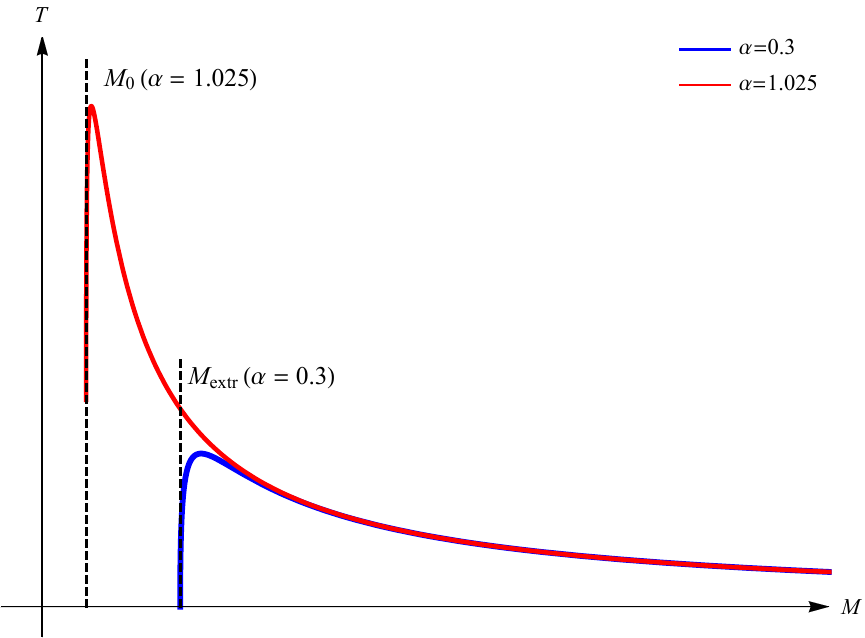}
    \caption{}
    \label{H}
\end{subfigure}
\hspace{-0.1cm}
\begin{subfigure}{0.48\linewidth}
    \centering
    \includegraphics[width=\linewidth]{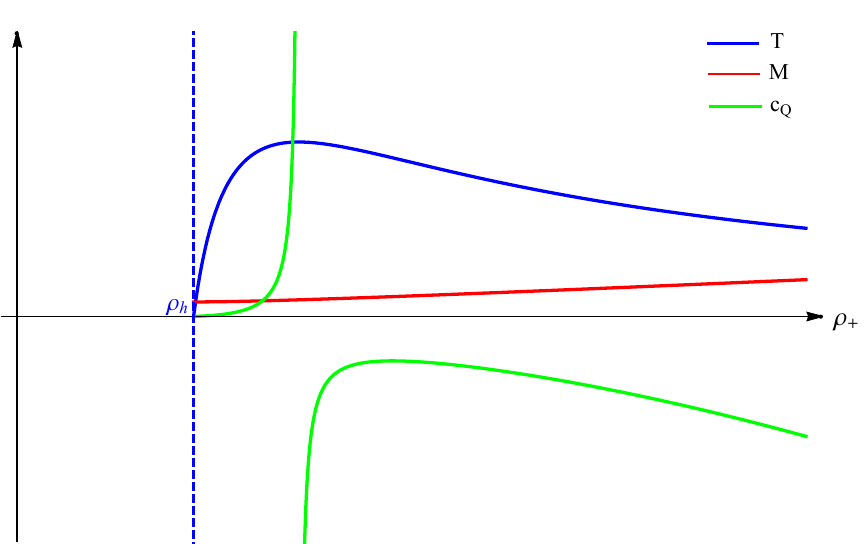}
    \caption{}
    \label{fig:T_M_cQ}
\end{subfigure}
\caption{\textbf{Left:} Parametric plot of the BH temperature $T$ as a function of the mass $M$, obtained from Eqs.~(\ref{T}) and (\ref{M}), for $\alpha=0.3$ (blue) and $\alpha=1.025$ (red). In the former case, the temperature vanishes at the extreme mass $M_{\rm extr}=C_{\rm extr} r_o/(2G)$, while in the latter case it reaches a finite value at $M_0=C_{\rm 0} r_o/(2G)$. \textbf{Right:} Plots of the temperature $T$ (Eq. \eqref{T}), the mass $M$ (Eq. \eqref{M}) and the specific heat $c_Q$ (Eq. \eqref{eq:specific_heat}) as functions of the event horizon radius $\rho_+$, for $\alpha=0.5$.  The blue dashed line indicates the extremal horizon radius $\rho_h$ given by Eq. \eqref{eq:extremal_horizon}.}
\end{figure}

The general relation $T$ vs.~$M$ can be visualized through a parametric plot, starting from Eqs.~(\ref{T}) and (\ref{M}). Some atypical feature should be expected in this anomalous case, since the behavior of $f(\rho)$ at the singularity drastically differs from what happens in the standard case. Indeed, while $f(\rho)$ in the standard case is divergent at the singularity at $\rho=0$ (see Eq.~\eqref{eq:c_crit}), in the anomalous case both $f(\rho)$ and its derivative are finite at the curvature singularity at $\rho=1$. As a consequence, there is a region of the space $(\alpha, C)$ where the temperature $T=0$ is not accesible, contrary to what would be expected for a charged BH. In this region, the curve $T$ vs.~$M$ does exhibit a maximum (it displays branches with positive and negative $c_Q$), but it remains truncated without reaching $T=0$. The inaccessibility of $T=0$ (i.e., of $f'(\rho_+)=0$) is marked by the condition $f(\rho=1)<0$ in Figure \ref{fig:c0}, or the condition $C>C_0(\alpha)$ in Eq.~\eqref{eq:c_0}, which is the region of a unique horizon.

Figure \ref{H} shows that the specific heat, \(c_Q=\partial M/\partial T\vert_\alpha\), is positive for small masses and negative for large masses, and diverges at the value of $M$ where the temperature reaches its maximum.  This behaviour is typical of charged BHs. The specific heat as a function of $\rho_+$ is 
\begin{equation} \label{eq:specific_heat}
    c_Q=\frac{\partial M}{\partial T}\Big\vert_\alpha=\frac{\frac{\partial M}{\partial \rho_+}}{\frac{\partial T}{\partial \rho_+}}=\frac{2\pi\, r_o^2}{G}\ \frac{\rho_+^2\,\sqrt{\rho_+^4-1}\ \left(\alpha-\rho_+^2+\sqrt{\rho_+^4-1}\right)}{1+\rho_+^4-(\alpha+\rho_+^2)\,\sqrt{\rho_+^4-1}} \ .
\end{equation}
The specific heat tends to zero for $\rho_+$ approaching $1$; in this limit the temperature is $T=(\alpha-1)/(4\pi\, \alpha\, r_o)$ and the mass is $M= C_0(\alpha)r_o/(2\, G)\, $. The behavior of these quantities as functions of the event horizon radius is shown in Figure ~\ref{fig:T_M_cQ}. Starting from the extremal configuration at $\rho_h$, the Hawking temperature increases from zero until it reaches a maximum and then decreases monotonically. By contrast, the mass grows monotonically with $\rho_+$. As a consequence, the specific heat changes sign through a divergence at the temperature maximum, separating a locally thermodynamically stable branch ($c_Q>0$) from an unstable one ($c_Q<0$).

The BH entropy $S$ results from integrating the first law of thermodynamics at constant charge: $dM\vert_\alpha=T\, dS$. It satisfies the Bekenstein-Hawking area law since
\begin{equation}
    S=\int\frac{dM\vert_\alpha}{T}=\int T(\rho_+)^{-1}\, \frac{\partial M}{\partial\rho_+}\ d\rho_+=\int\frac{2\pi\, r_o^2\rho_+}{G}\ d\rho_+=\frac{\pi\, r_+^2}{G}=\frac{A}{4\, G}\ .
\end{equation}
Entropy has a positive minimum since $r_+\ge r_o$ ($\rho_+\ge 1$). 
\bigskip

The geometry (\ref{eq solucion dionico imaginario}) has already been considered in the literature, since it is sourced by the standard BI energy-momentum tensor of a point-like charge. Its thermodynamics has also been studied, although in somewhat different contexts \cite{Dey,Cai,Fernando,Miskovic}. In this case, the results are as follows,
\begin{equation}\label{T2}
    T=\frac{1}{4\pi r_o}\, \frac{\alpha+\rho_+^2-\sqrt{\rho_+^4+1}}{\alpha\,\rho_+}\ ,
\end{equation}
\begin{equation}\label{M2}
    M=\frac{r_o}{6G\alpha}\, \left[\rho_+^3+3\alpha\rho_+ -\rho_+\sqrt{\rho_+^4+1}+2\ {_2 F_1}\left(\frac{1}{4},\frac{1}{2};\frac{5}{4};-\rho_+^{-4}\right)\right]\ .
\end{equation}
The graph $T$ vs.~$M$ exhibits similar characteristics than the previous one. The corresponding expressions for the specific heat and the entropy are
\begin{equation}
    c_Q=\frac{\partial M}{\partial T}\Big\vert_\alpha=\frac{2\pi\, r_o^2}{G}\ \frac{\rho_+^2\,\sqrt{\rho_+^4+1}\ \left(\alpha+\rho_+^2-\sqrt{\rho_+^4+1}\right)}{1-\rho_+^4-(\alpha-\rho_+^2)\,\sqrt{\rho_+^4+1}} \ , \ \ \ \ \ \ \ S=\frac{A}{4\, G}\ .
\end{equation}
As in the previous case, the entropy satisfies the Bekenstein--Hawking area law. The specific heat goes to zero for $\rho_+$ approaching $0$, whenever $\alpha\neq 1$; in this limit the temperature diverges and the mass is $M= C_{\text{crit}}(\alpha)r_o/(2\, G)\,$

Therefore, both solutions exhibit the standard thermodynamic behavior expected for charged black holes. The Hawking temperature vanishes in the extremal limit, the entropy satisfies the Bekenstein--Hawking area law, and the specific heat changes sign through a divergence that separates locally stable and unstable branches.

\subsection{The smoothing of singularities}
As mentioned previously, the choice of an imaginary $\beta$ in the
determinantal Lagrangian turns the picture $g$-electrodynamics into
its standard BI form. In fact, the electric field of the
point charge becomes regular at the origin; thus, the field
invariants $S$ and $P$ will not diverge at the origin as long as the
magnetic monopole $p$ is zero. Then, it is natural to wonder about
the impact of this regularity on the geometry of spacetime. First,
note that the condition $p=0$ has no impact on geometry because
$p$ enters the geometry in the combination $(q^2+p^2)^{1/2}$, since
that is the way $q$ and $p$ appear in the energy-momentum tensor
(\ref{Tanom dionico}) as a factor in $r_o$.  Second, the regular
behavior of the electromagnetic field does not prevent the geometry
from diverging. A direct inspection of  Eq.~(\ref{curvatura escalar
dionico}) shows that the curvature scalar $R$ diverges regardless of
whether $\beta$ is real or imaginary in the determinantal
Lagrangian. If $\beta$ is real (anomalous case), then $R$ diverges
at $\rho=1$ ($r=r_o$). If $\beta$ is imaginary (standard case), then
$R$ is obtained from Eq.~(\ref{curvatura escalar dionico}) by
replacing $\beta^2$ with $ -\beta^2$ (that is, $r_o^4\rightarrow
-r_o^4$) and $\epsilon$ with $-\epsilon$; thus, $R$ becomes regular
at $\rho=1$ but still diverges at $\rho=0$.

The behavior of the function $f(\rho)$ is not better in the standard
case than in the anomalous case. In fact, $f(\rho)$ is regular at
$\rho=1$ in the anomalous case (but its derivative diverges).
Instead, $f(\rho)$ diverges at $\rho=0$ in the standard case. Thus,
the way in which the Ricci tensor (\ref{eq Ricci dionico}) diverges
in each case is quite different. In the anomalous case, it results
(we use $\lambda=1$ to exclude the contribution of the cosmological
constant)
\begin{equation}  R^t_{\ t}=R^r_{\ r}=\frac{1}{2\epsilon\,\sqrt{\rho-1}}-\frac{1}{\epsilon}+\mathcal{O}(\sqrt{\rho-1})\ ,\ \ \ \ \ \  R^\theta_{\ \theta}=R^\phi_{\ \phi}=-\frac{1}{\epsilon}+\mathcal{O}(\sqrt{\rho-1})\ ,
\end{equation}
while in the standard case it is
\begin{equation}  R^t_{\ t}=R^r_{\ r}=\frac{1}{\epsilon}+\mathcal{O}(\rho^2)\ ,\ \ \ \ \ \  R^\theta_{\ \theta}=R^\phi_{\ \phi}=-\frac{1}{\epsilon\, \rho^2}+\frac{1}{\epsilon}+\mathcal{O}(\rho^2)\ .
\end{equation}
We can compare these results with the Ricci tensor in
Reissner-Nordstr\"{o}m geometry,
\begin{equation} \text{Reissner-Nordstr\"{o}m:}\ \ \ \ \ R^{t}_{\ t}=R^r_{\ r}=-R^\theta_{\ \theta}=-R^\phi_{\ \phi}=\frac{1}{2\epsilon\, \rho^4}\ .
\end{equation}
Therefore, BI electrogravity, both in its anomalous and
standard versions, alleviates the singular behavior of the geometry
at $\rho=1$ and $\rho=0$ respectively. We can confirm this property
at the level of the Riemann tensor by means of the Kretschmann
scalar $K = R_{\alpha \beta \gamma \delta} R^{\alpha \beta \gamma
\delta}$.  Its expression in terms of $f(\rho)$ and its derivatives is given by 
\begin{equation} \label{eq: expr_Kretschmann}
    K =   \frac{4\,[1-f(\rho)]^2+4\,\rho^2\, f'(\rho)^2 +\rho^4\, f''(\rho)^2}{r_o^4\ \rho^4}\ .
\end{equation}
In the anomalous case (\ref{eq solucion dionico}), the behavior of
$K$ at $\rho=1$ is
\begin{equation}\label{K dionico}
    K=\frac{1}{\epsilon^2\,(\rho-1)}-\frac{4}{3}\,\frac{1+3\,\alpha\, C-2 \sqrt{\pi}\ \frac{\Gamma(\frac{5}{4})}{\Gamma(\frac{3}{4})}}{\epsilon^2\,\sqrt{\rho-1}}+\mathcal{O}(1)\ ;
\end{equation}
while in the standard case (\ref{eq solucion dionico imaginario}),
the behavior of $K$ at $\rho=0$ is
\begin{equation}~\label{K dioimaginario}
    K=\frac{4}{3\pi}\,\frac{\left(3\sqrt{\pi}\, \alpha\, C-2\, \Gamma(\frac{1}{4}) \Gamma(\frac{5}{4})\right)^2}{\epsilon^2\,\rho^6}+\mathcal{O}(\rho^{-5})\ .
\end{equation}
On the other hand, the Kretschmann scalar for Reissner-Nordstr\"{o}m
geometry is
\begin{equation} \text{Reissner-Nordstr\"{o}m:}\ \ \ \ \ K=2\ \frac{1+6\,(1-\alpha\, C\, \rho)^2}{\epsilon^2\ \rho^8}=\frac{8G^2Q^4}{r^8}\,\left[1+6\left(\frac{Mr}{Q^2}-1\right)^2\right]\ .
\end{equation}
This means that the ability to smooth the geometry is better in the
anomalous case (when $\beta$ is real in the determinantal
Lagrangian), because the singularity has been shifted to $\rho=1$,
on a sphere of area $4\pi\,r_o^2$; thus the pathological effect of
the factor $\rho^{-4}$ in Eq.~(\ref{eq: expr_Kretschmann}) is
circumvented.

The divergent behavior of $K$ in the standard case of Eq.~(\ref{K
dioimaginario}) is also better than in the Reissner-Nordstr\"{o}m
geometry, since $K$ maintains the behavior $\rho^{-6}$ characteristic of
the (chargeless) Schwarzschild geometry. Noticeably, this behavior
can still be improved by choosing a mass such that \footnote{This
mass is extremal for $\alpha=1$ (see Eq.~(\ref{eq C
dioimaginario})).}
\begin{equation}\label{choice}
    C=C_{\text{crit}}=\frac{2\ \Gamma(\frac{1}{4})\Gamma(\frac{5}{4})}{3\sqrt{\pi}\, \alpha}\ .
\end{equation}
This happens because this mass value regularizes the function
$f(\rho)$ at the origin, since $f(\rho)$ in Eq.~(\ref{eq solucion
dionico imaginario}) turns to be
\begin{equation}
    f(\rho)=\frac{\alpha-1}{\alpha}+\frac{\rho^2}{3\,\alpha}+\mathcal{O}(\rho^4)
\end{equation}
(a deficit angle is formed at the origin). With the choice
(\ref{choice}), the Kretschmann scalar of the standard case (\ref{eq
solucion dionico imaginario}) becomes 
\begin{equation}
      K=\frac{4}{\epsilon^2\rho^4}-\frac{8}{3\epsilon^2\rho^2}+\frac{52}{15\,\epsilon^2}+\mathcal{O}(\rho^{2})\ .
\end{equation}
Instead, the behavior of $K$ in the anomalous case cannot be
improved by choosing a particular value for $C$, since the first
term in Eq.~(\ref{K dionico}) does not depend on $C$.  Neither the
behavior of $K$ in Reissner-Nordstr\"{o}m geometry could be improved in
this way. Consistently, the same $r^{-4}$ dependence of the
Kretschmann scalar was found in \cite{yang} solving the Einstein
equations sourced by the BI electromagnetism, for a dyonic
BH and a critical value of the mass.

The existence of a true singularity is confirmed by the
fact that both $\rho=1$ in the anomalous case, or $\rho=0$ in the
standard case can be reached in a finite proper time. This is
because the behavior of $f$ at the singularity is either regular
(anomalous case) or the same as in an ordinary BH (standard
case). Therefore, spacetime is geodesically incomplete.

\section{Conclusions} \label{sec:concl}
Born--Infeld electrogravity is governed by a Lagrangian that couples gravity and electromagnetism through a unique determinantal structure, in which the Lagrangian density is given by the determinant of a tensor whose symmetric part is geometric and whose antisymmetric part is electromagnetic. The Palatini formalism extracts the dynamics by taking the metric \(\bf g\), the affine connection \(\bf \Gamma\), and the electromagnetic potential \(\bf A\) as independent dynamical variables. It exhibits its full power by yielding well-behaved dynamic equations free of instabilities. The variation with respect to \(\bf g\) leads to a constraint equation, which is the consequence of the absence of derivatives of the metric in the Lagrangian. In turn, the dynamical content of the theory comes from the variations with respect to \(\bf \Gamma\) and \(\bf A\). The variation with respect to \(\bf \Gamma\) must  preserve the symmetry attributed to the geometric sector; this is achieved by setting the torsion to zero. This condition, together with the constraint equation, leads to the metric compatibility and fixes the connection as the Levi-Civita connection.

 As they arise from the Palatini formalism, the field equations are separated in terms of the symmetric and antisymmetric parts of the auxiliary tensor \(\bar q^{\mu\nu}\). The dynamics of \(\bar q^{[\mu\nu]}\) is the standard Born--Infeld electrodynamics as developed over an effective geometric background defined by the symmetric tensor \(\mathcal
G_{\mu\nu}\equiv g_{\mu\nu}+\epsilon R_{(\mu\nu)}=q_{(\mu\nu)}\). Nevertheless, an electrodynamical picture in the background of the physical metric  \(g_{\mu\nu}\) is also realizable by combining the symmetric and antisymmetric sectors of the dynamics. This second
description retains its Born--Infeld essence but exhibits a sign alteration in two terms that contain the quadratic invariants $S$ and $P$;
therefore, we refer to it as ``anomalous''. We have shown that these two pictures, ``\(\mathcal{G}\)'' and ``\(g\)'', are related by a
set of identities that map the pseudoscalars built with \(\mathcal
G\) into those built with \(g\), and that also relate the
corresponding volume elements (Eqs.~\eqref{PPGg}, \eqref{S 2} and \eqref{conversion}).
In particular, these relations explain how the apparent coupling of
the electromagnetic field to curvature in picture \(\mathcal{G}\)
(Eq.~\eqref{eq em 0}) is reabsorbed and translated into a minimally
coupled Born--Infeld structure in picture \(g\), in agreement with
the equivalence principle. It is worth noting that the anomalous and standard characters of each picture are not essential since they can be interchanged by replacing $(\beta,\epsilon)$ with $(i\beta,-\epsilon)$,  which is allowed
because only even powers of $\beta$ appear in the Lagrangian. These two electrodynamical pictures, in the backgrounds \(\mathcal{G}\) and \(g\), resemble Pleba\'nski's developments \cite{Tmunu_BI2} that show that the same field configuration can be conceived in two different metrics related through the energy-momentum tensor of the field (see also \cite{Tmunu_BI4}). In our case this is so, because \(\epsilon R_{\mu\nu}\) in \(\mathcal{G}_{\mu\nu}\) are connected to the electromagnetic energy-momentum tensor through the Einstein equations.

To explore the implications of the model on the singular behavior of
the solutions, we analyzed spherically symmetric dyonic
configurations. If $\beta$ is real in the Lagrangian, then the electric field becomes singular at a
finite radius \(r_o\), restricting the domain of the solution to
\(r>r_o\), while the magnetic field maintains its Coulombian form. If
instead the parameter $\beta$ is taken to be imaginary, the
electric field acquires the standard Born--Infeld form but the magnetic
monopole is divergent at the origin. Remarkably, even though the
electric and magnetic monopoles enter the field invariants $S$ and
$P$ unequally, they contribute to the energy and pressure on an
equal footing through the combination \(\sqrt{q^2+p^2}\), a feature
that is inherited by the geometry. The obtained geometric solution
approaches the Reissner--Nordstr\"{o}m form asymptotically, in the weak gravity region. The family of
solutions admits one, two or no horizons depending on the
mass and parameter choices; their thermodynamic properties are governed by the behavior of the function $f(\rho)$, whether of Schwarzschild type or RN type.

The relation mass vs.~charge for the extremal BH is similar to the one of the extremal Reissner--Nordstr\"{o}m BH if the charge is large ($\alpha<1$) but differs significantly for small charge, because the mass of the BI extremal BH goes to $M_{\text{extr}}=\beta\, c^4/(4G^{3/2})$ for decreasing charge (increasing values of $\alpha>1$). In this limit, the mass and the horizon radious are characterized exclusively by the fundamental constants of the theory. The horizon radious of this fundamental extremal BH satisfies $r_h=2GM_{\text{extr}}$, like the Schwarzschild BH. However, unlike the Schwarzschild BH, its mass is not an integration constant but the natural unit of mass defined by the fundamental constants of the theory; besides, its temperature is zero, as is typical of an extremal BH.

With regard to the singularities, Born--Infeld electrogravity alleviates the
geometric divergences compared to the Reissner--Nordstr\"{o}m case, but
nevertheless spacetime remains geodesically incomplete. The
improvement is more pronounced in the real-$\beta$ case,
because the curvature singularity is shifted from the center to a
sphere at \(r=r_o\), mitigating the pathological inverse power
behavior of the Kretschmann scalar. Noticeably, for imaginary $\beta$
there exists a solution for which the metric becomes regular at the
origin and the singularity in the Kretschmann scalar is of order
$r^{-4}$.

Finally, it would be interesting to investigate the cosmological solutions that arise from the here considered Lagrangian. In particular, in \cite{banados}, the authors analyzed a type of BI gravity and found that the corresponding cosmological solution is free from singularities.

\bigskip

\textbf{Acknowledgments:} The authors were supported by Consejo Nacional de Investigaciones Cient\'{\i}ficas y T\'{e}cnicas (CONICET) and Universidad de Buenos Aires (UBA).

\appendix

\section{Determinant of $q$}

\label{apendice:desarrollo_q}

The determinant of $q$ is the essential piece of Lagrangian (\ref{eq
accion}). Since $q=\mathcal{G}+\beta ~F=\mathcal{G}\ (I+\beta
~\breve{F})$, then
\begin{equation}
\det q=\det \mathcal{G~}\det (I+\beta ~\breve{F})
\end{equation}
To compute $\det (I+\breve{F})$ we will use an expansion that is valid for any matrix $M$ in an arbitrary dimension $n$. Let us call $%
s_{k}$ the traces of the powers of $M$,
\begin{equation}
s_{k}=\operatorname{Tr}(M^{k}),\quad 1\leq k\leq n\ .
\end{equation}
Then the determinant of $I-M$ can be expressed as
\begin{equation}
\det (I-M)=1+\sum_{k=1}^{n}p_{k},  \label{desarrollo_det}
\end{equation}
where the coefficients $p_{k}$ are related to the traces $s_{k}$ as
follows
\begin{equation}
p_{1} =-s_{1}, \quad
p_{2} =-\frac{1}{2}(s_{2}+p_{1}s_{1}), \quad
p_{3} =-\frac{1}{3}(s_{3}+p_{1}s_{2}+p_{2}s_{1}), \quad
p_{4} =-\frac{1}{4}(s_{4}+p_{1}s_{3}+p_{2}s_{2}+p_{3}s_{1}), \quad
...
\label{p's}
\end{equation}
In our case it is $n=4$, and $M=-\beta ~\breve{F}$. Then,
\begin{equation}
s_{1}=\operatorname{Tr}(M)=-\beta
~\operatorname{Tr}(\breve{F})=-\beta
~\operatorname{Tr}(\mathcal{\bar{G}}F)=0~,
\end{equation}
since $\mathcal{\bar{G}}=\mathcal{G}^{-1}$ is symmetric but $F$ is
antisymmetric. Besides \footnote{$s_3$ is zero because
${F}_{\lambda\rho }$ and $\breve{F}^{\rho\nu }$ are antisymmetric.
In fact, it results that $\breve{F}_{~~\lambda }^{\nu
}\breve{F}_{~~\rho }^{\lambda }\breve{F}_{~~\nu }^{\rho
}=\breve{F}^{\nu\lambda }{F}_{\lambda\rho }\breve{F}^{\rho\nu
}=-\breve{F}^{\lambda\nu }{F}_{\rho\lambda }\breve{F}^{\nu\rho
}=-\breve{F}_{~~\lambda }^{\nu }\breve{F}_{~~\rho }^{\lambda
}\breve{F}_{~~\nu }^{\rho }$.}
\begin{eqnarray}
s_{2}&=&\operatorname{Tr}(MM)=\beta ^{2}~\breve{F}_{~~\lambda }^{\nu
}\breve{F}_{~~\nu
}^{\lambda }=-4\beta ^{2}\breve{S}~,\\ \notag\\
s_{3}&=&\operatorname{Tr}(MMM)=-\beta ^{3}~\breve{F}_{~~\lambda
}^{\nu }\breve{F}_{~~\rho
}^{\lambda }\breve{F}_{~~\nu }^{\rho }=0\ ,\\ \notag\\
s_{4} &=&\operatorname{Tr}(MMMM)=\beta ^{4}~\breve{F}_{~~\lambda
}^{\nu }\breve{F}_{~~\rho
}^{\lambda }\breve{F}_{~~\eta }^{\rho }\breve{F}_{~~\nu }^{\eta }=\beta ^{4}~%
\breve{F}_{~~\lambda }^{\nu }(-2\breve{S}~\delta _{\eta }^{\lambda }+^{\ast }%
\breve{F}_{~\rho }^{\lambda }~^{\ast }\breve{F}_{\eta }^{\rho })~\breve{F}%
_{~~\nu }^{\eta }~\notag \\
&=&\beta ^{4}~(8~\breve{S}^{2}+\breve{P}^{2}~\delta _{\rho }^{\nu
}\delta _{\nu }^{\rho })=4\beta ^{4}~(2~\breve{S}^{2}+\breve{P}^{2})
\end{eqnarray}
So it is
\begin{equation}
p_{1}=0~,~~~~~p_{2}=-\frac{s_{2}}{2}=2\beta ^{2}\breve{S}%
~,~~~~~p_{3}=0~,~~~~~p_{4}=-\frac{1}{4}(s_{4}+p_{2}s_{2})=-\beta ^{4}~\breve{%
P}^{2}~.
\end{equation}
Therefore
\begin{equation}
\det q=\det \mathcal{G~}\det (I+\beta ~\breve{F})=(\det \mathcal{G)~}%
(1+2\beta ^{2}\breve{S}-\beta ^{4}~\breve{P}^{2})~.  \label{detq}
\end{equation}
Let us mention that the same treatment could be given to
$\det\mathcal{G}=\det g\, \det(\delta^\mu_\nu+\epsilon R^\mu_{\
\nu})$, or even to $\det q=\det g\, \det(\delta^\mu_\nu+\epsilon
R^\mu_{\ \nu}+\beta F^\mu_{\ \nu})$. However, the traces of
$R^\mu_{\ \nu}$ and its powers are all different from zero, which
leads to a much more complicated expression.

\section{The projective mode} \label{app:Projective}
In Cartan language, curvature is expressed as a family of 2-forms,
\begin{equation} \label{riemann}
\widetilde{\widetilde{\mathcal{R}}}{}^a_{\;b} \equiv
d\tilde{\omega}^a_{\;b} + \tilde{\omega}^a_{\;c} \wedge
\tilde{\omega}^c_{\;b}\ ,
\end{equation}
labeled by two tangent space indices. The family of 1-forms
$\tilde{\omega}^a_{\;b}$ constitutes the \textit{spin connection},
defined as
\begin{equation}
\tilde{\omega}^a_{\;b} = \Gamma^a_{\;bc} {\tilde e}^c,
\end{equation}
where $\{ \tilde{e}^c \}$ is the co-tangent space basis, dual to the
tangent basis $\{ \vec{e}_c \}$, and $\Gamma^a_{bc}$ are the
components of the affine connection in that basis
($\nabla_{\vec{e}_c} \vec{e}_b = \Gamma^a_{bc} \, \vec{e}_a$). The
usual components of the Riemann tensor $R^a{}_{bcd}$ then
satisfy\footnote{Coordinate bases are used in the main body of this
article: $\{\tilde{e}^c\} \rightarrow \{dx^\mu\}$.}
\begin{equation} \label{riemann1}
\widetilde{\widetilde{\mathcal{R}}}{}^a_{\;b}=\frac{1}{2} \, R^a_{\,
bcd}\; \tilde{e}^c\wedge \tilde{e}^d\ .
\end{equation}
The antisymmetry in the last two indices of $R^a{}_{bcd}$ reflects
the 2-form nature of
$\widetilde{\widetilde{\mathcal{R}}}{}^a_{\;b}$, while the indices
$a$, $b$ label the different members of the family.
Let us now consider a \textit{projective transformation} of the
connection
\begin{equation}
\tilde{\omega}^a_{\;b} \rightarrow \tilde{\omega}^a_{\;b} +
\delta^a_{b}\; \tilde{\mathcal{A}}\ ,
\end{equation}
where $\tilde{\mathcal{A}} = \mathcal{A}_c \, \tilde{e}^c$ is an
arbitrary 1-form. This transformation is equivalent
to\footnote{Under this transformation, the covariant derivative of
vector $\bar{U}$ in the direction of vector $\bar{V}$ acquires a
term proportional to $\bar{U}$: $\nabla_{\bar{V}} \bar{U}
\rightarrow \nabla_{\bar{V}} \bar{U} + \mathcal{A}_c V^c \,
\bar{U}$. Along an autoparallel, where $\nabla_{\bar{U}} \bar{U} =
0$, the term can be absorbed via a redefinition of the affine
parameter.}
\begin{equation}
    \Gamma^a_{bc}\longrightarrow  \Gamma^a_{bc}+\delta^a_b\, \mathcal{A}_c\ .
\end{equation}
Under this transformation, the curvature becomes
\begin{equation} \label{riemann_mod}\widetilde{\widetilde{\mathcal{R}}}{}^a_{\;b} \rightarrow d\tilde\omega^a_{\;b}+ \delta^a_{b}\; d\tilde{\mathcal{A}}   + \left(\tilde\omega^a_{\;c} + \delta^a_{c} \;\tilde{\mathcal{A}}\right) \wedge \left(\tilde\omega^c_{\;b} + \delta^c_{b} \;\tilde{\mathcal{A}} \right)\ ,
\end{equation}
which, using the antisymmetry of the wedge product, simplifies to
\begin{equation} \widetilde{\widetilde{\mathcal{R}}}{}^a_{\;b}\rightarrow \widetilde{\widetilde{\mathcal{R}}}{}^a_{\;b} + \delta^a_{b} \; d\tilde{\mathcal A} \ ,
\end{equation}
or, in a coordinate basis,
\begin{equation}
R^\lambda_{\;\rho\mu\nu} \rightarrow R^\lambda_{\;\rho\mu\nu} +
\delta^\lambda_{\rho} \,(\partial_\mu {\mathcal A}_\nu -
\partial_\nu {\mathcal A}_\mu)\ .
\end{equation}
Consequently, the Ricci tensor transforms as
\begin{equation}
R_{\rho\nu} \rightarrow R_{\rho\nu} + (\partial_\rho {\mathcal
A}_\nu - \partial_\nu {\mathcal A}_\rho)\ .
\end{equation}
Even if the Ricci tensor was initially symmetric, the projective
transformation breaks that symmetry. However, in a Lagrangian of the
form (\ref{eq accion})--(\ref{eq q}), a variation of the connection
of the type $\delta\Gamma^\alpha_{\mu\nu}=\delta^\alpha_\mu\,
\delta{\mathcal A}_\nu$ is indistinguishable from a variation of the
electromagnetic potential $\delta A_\nu = \epsilon \beta^{-1}
\delta\mathcal{A}_\nu$. This means that, in a Lagrangian like the
one considered here, the projective degrees of freedom are
degenerate with the electromagnetic ones. Freezing the projective
variations of the connection (thus preserving the symmetry of the
Ricci tensor) and retaining only variations of the electromagnetic
potential is merely a gauge fixing that resolves this ambiguity.


\begin{thebibliography}{99}

\bibitem{born-33} M. Born, ``Modified field equations with a finite radius of the electron'', Nature \textbf{132}, 282 (1933). \href{https://doi.org/10.1038/132282a0}{doi:10.1038/132282a0}

\bibitem{borninfeld-34-Nature} M. Born and L. Infeld, ``Foundations of the new field theory'', Nature \textbf{132}, 1004 (1933). \href{https://doi.org/10.1038/1321004b0}{doi:10.1038/1321004b0}

\bibitem{born-34} M. Born, ``On the quantum theory of the electromagnetic field'', Proc. R. Soc. (London) A  \textbf{143}, 410-437 (1934). \href{https://doi.org/10.1098/rspa.1934.0010}{doi:10.1098/rspa.1934.0010}

\bibitem{borninfeld-34} M. Born and L. Infeld, ``Foundations of the new field theory'', Proc. R. Soc. (London)  A \textbf{144}, 425-451 (1934). \href{https://doi.org/10.1098/rspa.1934.0059}{doi:10.1098/rspa.1934.0059}

\bibitem{deser} S. Deser and G. W. Gibbons, ``Born-Infeld-Einstein actions?'',  Class. Quantum Grav. \textbf{15}, 35 (1998). \href{https://doi.org/10.1088/0264-9381/15/5/001}{doi:10.1088/0264-9381/15/5/001}

\bibitem{freigen} J. A. Feigenbaum, P. O. Freund, and M. Pigli, ``Gravitational analogues of nonlinear Born electrodynamics'', Phys. Rev. D \textbf{57}, 4738 (1998). \href{https://doi.org/10.1103/PhysRevD.57.4738}{doi:10.1103/PhysRevD.57.4738}. \href{https://arxiv.org/abs/hep-th/9709196}{arXiv:hep-th/9709196}

\bibitem{freigen4D} J. A. Feigenbaum, ``Born-regulated gravity in four dimensions'', Phys. Rev. D \textbf{58}, 124023 (1998).  \href{https://doi.org/10.1103/PhysRevD.58.124023}{doi.org/10.1103/PhysRevD.58.124023}. \href{https://arxiv.org/abs/hep-th/9807114}{arXiv:hep-th/9807114}

\bibitem{vollick04} D. N. Vollick, ``Palatini approach to Born-Infeld-Einstein theory and a geometric description of electrodynamics'', Phys. Rev. D \textbf{69}, 064030 (2004). \href{https://doi.org/10.1103/PhysRevD.69.064030}{doi:10.1103/PhysRevD.69.064030}. \href{https://arxiv.org/abs/gr-qc/0309101}{arXiv:gr-qc/0309101}

\bibitem{banados} M. Ba\~nados and P. G. Ferreira, ``Eddington's theory of gravity and its progeny'', Phys. Rev. Lett. \textbf{105}, 011101  (2010).  \href{https://doi.org/10.1103/PhysRevLett.105.011101}{doi:10.1103/PhysRevLett.105.011101}. \href{https://arxiv.org/abs/1006.1769}{arXiv:1006.1769}. [erratum: Phys. Rev. Lett. \textbf{113}, 119901 (2014). \href{https://doi.org/10.1103/PhysRevLett.113.119901}{doi:10.1103/PhysRevLett.113.119901}].

\bibitem{review_beltran} J. Beltr\'{a}n Jim\'{e}nez, L. Heisenberg, G. J. Olmo and D. Rubiera-Garcia, ``Born-Infeld inspired modifications of gravity'', Phys. Rept. \textbf{727}, 1-129 (2018).
\href{https://doi.org/10.1016/j.physrep.2017.11.001}{doi:10.1016/j.physrep.2017.11.001}.
\href{https://arxiv.org/abs/1704.03351}{arXiv:1704.03351}

\bibitem{MTG} C. M. Will, ``Theory and Experiment in Gravitational Physics'', Cambridge University Press, New York (1993).

\bibitem{Einstein_equivalence} C. M. Will, ``The confrontation between general relativity and experiment'', Living Rev. Rel. \textbf{17}, 4 (2014). \href{https://doi:10.12942/lrr-2014-4}{doi:10.12942/lrr-2014-4}. \href{https://arxiv.org/abs/1403.7377}{arXiv:1403.7377}

\bibitem{ambiguedades} A. Delhom, ``Minimal coupling in the presence of non-metricity and torsion'',  Eur. Phys. J. C  \textbf{80}, 728 (2020). \href{https://doi.org/10.1140/epjc/s10052-020-8330-y}{doi:10.1140/epjc/s10052-020-8330-y}. \href{https://arxiv.org/abs/2002.02404}{arXiv:2002.02404}

\bibitem{Vollick05} D. N. Vollick, ``Born-Infeld-Einstein theory with matter'', Phys. Rev. D \textbf{72}, 084026 (2005). \href{https://doi.org/10.1103/PhysRevD.72.084026}{doi:10.1103/PhysRevD.72.084026}. \href{https://arxiv.org/abs/gr-qc/0506091}{arXiv:gr-qc/0506091}

\bibitem{afonso_et_al} V. I. Afonso, C. Bejarano, R. Ferraro and G. J. Olmo, ``Determinantal Born-Infeld Coupling of Gravity and Electromagnetism'', Phys. Rev. D \textbf{105}, 084067 (2022). \href{https://doi.org/10.1103/PhysRevD.105.084067}{doi:10.1103/PhysRevD.105.084067}. \href{https://arxiv.org/abs/2112.09978}{arXiv:2112.09978}

\bibitem{Exirifard}Q. Exirifard and M. M. Sheikh-Jabbari, ``Lovelock gravity at the crossroads of Palatini and metric formulations'', Phys. Lett. B \textbf{661}, 158 (2008). \href{https://doi.org/10.1016/j.physletb.2008.02.012}{doi:10.1016/j.physletb.2008.02.012}. \href{https://arxiv.org/abs/0705.1879}{arXiv:hep-th/0705.1879}


\bibitem{Sotiriou} T.P. Sotiriou and V. Faraoni, ``\textit{f(R)} theories of gravity'',
Rev. Mod. Phys. \textbf{82}, 451 (2010). \href{ https://doi.org/10.1103/RevModPhys.82.451}{doi:10.1103/RevModPhys.82.451}. \href{https://arxiv.org/abs/0805.1726}{arXiv:gr-qc/0805.1726}

\bibitem{Palatini} G. J. Olmo, ``Palatini Approach to Modified Gravity: f(R) Theories and Beyond'', Int. J. Mod. Phys. D \textbf{20}, 413 (2011). \href{https://doi.org/10.1142/S0218271811018925} {doi:10.1142/S0218271811018925}. \href{https://arxiv.org/abs/1101.3864}{arXiv:1101.3864}

\bibitem{Galtsov} D. Gal'tsov and S. Zhidkova,
``Ghost-free Palatini derivative scalar–tensor theory: Desingularization and the speed test'', Phys. Lett. B, \textbf{790}, 453 (2019). \href{https://doi.org/10.1016/j.physletb.2019.01.061}{doi:10.1016/j.physletb.2019.01.061}. \href{https://arxiv.org/abs/1808.00492}{arXiv:1808.00492}

\bibitem{Aoki} K. Aoki and K. Shimada, ``Scalar-metric-affine theories: Can we get ghost-free theories from symmetry?'', Phys. Rev. D \textbf{100}, 044037 (2019). \href{https://doi.org/10.1103/PhysRevD.100.044037}{10.1103/PhysRevD.100.044037}. \href{https://arxiv.org/abs/1904.10175}{arXiv:1904.10175} 



\bibitem{Faraoni} N. Lanahan-Tremblay and V. Faraoni, ``The Cauchy problem of \textit{f(R)} gravity'', Class. Quantum Grav. \textbf{24}, 5667 (2007). \href{https://doi.org/10.1088/0264-9381/24/22/024}{doi:10.1088/0264-9381/24/22/024}, \href{https://arxiv.org/abs/0709.4414}{arXiv:gr-qc/0709.4414}

V. Faraoni, ``Reply to `A comment on `The Cauchy problem of \textit{f(R)} gravity' ' '', Class. Quantum Grav. \textbf{26}, 168002 (2009). \href{https://doi.org/10.1088/0264-9381/26/16/168002}{doi:10.1088/0264-9381/26/16/168002}. \href{https://arxiv.org/abs/0906.5311}{arXiv:gr-qc/0906.5311}

\bibitem{rbg1}
V. I. Afonso, G. J. Olmo and D. Rubiera-Garcia, ``Mapping
Ricci-based theories of gravity into general relativity'', Phys.
Rev. D \textbf{97}, 021503(R) (2018),
\href{https://doi.org/10.1103/PhysRevD.97.021503}
{doi:10.1103/PhysRevD.97.021503}
\href{https://arxiv.org/abs/1801.10406}{arXiv:1801.10406}
\bibitem{rbg2} V. I. Afonso, G. J. Olmo, E. Orazi and D. Rubiera-Garc\'{\i}a, ``Mapping nonlinear gravity into General Relativity with nonlinear electrodynamics'', Eur. Phys. J. C \textbf{78} 866 (2018). \href{https://doi.org/10.1140/epjc/s10052-018-6356-1}{doi:10.1140/epjc/s10052-018-6356-1}. \href{https://arxiv.org/abs/1807.06385}{arXiv:1807.06385}


\bibitem{Tmunu_BI2} J. Pleba\'nski, Lectures on non linear electrodynamics, Nordita Lecture Notes, Copenhagen (1968).

\bibitem{Tmunu_BI3} G. Boillat, ``Nonlinear Electrodynamics: Lagrangians and Equations of Motion'', J. Math. Phys. \textbf{11}, 941 (1970). \href{https://doi.org/10.1063/1.1665231}{doi:10.1063/1.1665231}

\bibitem{Tmunu_BI4} M. Novello, V. A. De Lorenci, J. M. Salim, and R. Klippert, ``Geometrical aspects of light propagation in nonlinear electrodynamics'', Phys. Rev. D \textbf{61} 045001 (2000). \href{https://doi.org/10.1103/PhysRevD.61.045001}{doi:10.1103/PhysRevD.61.045001}. \href{https://arxiv.org/abs/gr-qc/9911085}{arXiv:gr-qc/9911085}

\bibitem{desarrollo_Tmunu_BI} R. Ferraro, ``Born-Infeld electrostatics in the complex plane'', J. High Energ. Phys. 2010, \textbf{28} (2010). \href{https://doi.org/10.1007/JHEP12(2010)028}{doi:10.1007/JHEP12(2010)028}. \href{https://arxiv.org/abs/1007.2651}{arXiv:1007.2651}

\bibitem{projective2}
J. Beltr\'an Jim\'enez and A. Delhom, ``Instabilities in
Metric-Affine Theories of Gravity'',
 Eur. Phys. J. C \textbf{80} 6-585 (2020).
\href{https://doi.org/10.1140/epjc/s10052-020-8143-z}{doi:10.1140/epjc/s10052-020-8143-z}.
\href{https://arxiv.org/abs/2004.11357}{arXiv:2004.11357}

\bibitem{invarianciaproyectiva} V. I.  Afonso, C. Bejarano, J.  Beltr\'{a}n Jim\'{e}nez, G. J. Olmo, and E. Orazi, ``The trivial role of torsion in projective invariant theories of gravity with non-minimally coupled matter fields'', Class. Quantum Grav. \textbf{34} 235003 (2017). \href{https://iopscience.iop.org/article/10.1088/1361-6382/aa9151}{doi:10.1088/1361-6382/aa9151}. \href{https://arxiv.org/abs/1705.03806v2}{arXiv:1705.03806}

\bibitem{vollick06} D. N. Vollick, ``Black hole and cosmological space-times in Born-Infeld-Einstein theory'' (2006), arXiv:gr-qc/0601136. \href{https://arxiv.org/abs/gr-qc/0601136}
{arXiv:gr-qc/0601136}

\bibitem{Salazar} A. Garc\'{\i}a D., H. Salazar I. and  J. F. Pleba\'nski, ``Type-D solutions of the Einstein and Born-Infeld nonlinear-electrodynamics equations'', Nuovo Cimento B \textbf{84}, 65-90 (1984). \href{https://doi.org/10.1007/BF02721649}{doi:10.1007/BF02721649}

\bibitem{Breton} N. Bret\'{o}n, ``Geodesic structure of the Born-Infeld black hole'', Class. Quantum Grav. \textbf{19}, 601-612 (2002). \href{https://doi.org/10.1088/0264-9381/19/4/301}{doi:10.1088/0264-9381/19/4/301}

\bibitem{PoissonIsrael90} Poisson, E. and Israel, W., `` Internal structure of black holes '', Phys. Rev. D \textbf{41}, 1796-1809 (1990). \href{https://doi.org/10.1103/PhysRevD.41.1796}{doi:10.1103/PhysRevD.41.1796}

\bibitem{hale} T. Hale, R.A. Hennigar, and D. Kubiz\ifmmode \check{n}\else \v{n}\fi{}\'ak, ``Excising Cauchy Horizons with Nonlinear Electrodynamics'', Phys. Rev. D \textbf{113}, L061502 (2026). \href{https://doi.org/10.1103/x8c8-j85k} {doi:10.1103/x8c8-j85k}. \href{https://arxiv.org/abs/2506.20802 } {arXiv:2506.20802}

\bibitem{oliveira} H. P. de Oliveira, ``Non-linear charged black holes'', Class. Quantum Grav. \textbf{11}, 1469-1482 (1994).
\href{https://doi.org/10.1088/0264-9381/11/6/012}{doi:10.1088/0264-9381/11/6/012}

\bibitem{breton01} N. Bret\'{o}n, ``Born-Infeld generalization of the Reissner-Nordstrom black hole'' (2001), arXiv:gr-qc/0109022. \href{https://arxiv.org/abs/gr-qc/0109022}
{arXiv:gr-qc/0109022}

\bibitem{russo} J. G. Russo and P. K. Townsend, ``Black holes and causal nonlinear electrodynamics''. \href{https://arxiv.org/abs/2601.07789} {arXiv:2601.07789}

\bibitem{Ortin} T. Ortín,  ``Gravity and Strings'', Cambridge Monographs in Mathematical Physics,  Cambridge University Press,  Cambridge (2004). 

\bibitem{Dey} T. K. Dey, ``Born-Infeld black holes in the presence of a cosmological constant'', Phys. Lett. B \textbf{595}, 484-490 (2004). \href{https://doi.org/10.1016/j.physletb.2004.06.047}{doi:10.1016/j.physletb.2004.06.047}. \href{https://arxiv.org/pdf/hep-th/0406169}{arXiv:hep-th/0406169}  

\bibitem{Cai} R. G. Cai, D. W. Pang, and A. Wang, ``Born-Infeld Black Holes in (A)dS Spaces'', Phys. Rev. D \textbf{70}, 124034 (2004). \href{https://doi.org/10.1103/PhysRevD.70.124034}{doi:10.1103/PhysRevD.70.124034}. \href{https://arxiv.org/abs/hep-th/0410158}{arXiv:hep-th/0410158}

\bibitem{Fernando} S. Fernando, ``Thermodynamics of Born-Infeld-anti-de Sitter black holes in the grand canonical ensemble'', Phys.Rev. D \textbf{74}, 104032 (2006). \href{https://doi.org/10.1103/PhysRevD.74.104032}{doi:10.1103/PhysRevD.74.104032}. \href{https://arxiv.org/pdf/hep-th/0608040}{arXiv:hep-th/0608040}

\bibitem{Miskovic} O. Miskovi\'{c} and R. Olea, ``Thermodynamics of Einstein-Born-Infeld black holes with negative cosmological constant'', Phys. Rev. D \textbf{77}, 124048 (2008).
\href{https://doi.org/10.1103/PhysRevD.77.124048}{doi:10.1103/PhysRevD.77.124048}. \href{https://arxiv.org/abs/0802.2081}{arXiv:0802.2081} 

\bibitem{yang} Y. Yang, ``Dyonically Charged Black Holes Arising in Generalized Born--Infeld Theory of Electromagnetism'', Ann. Phys. \textbf{443}, 168996 (2022). \href{https://doi.org/10.1016/j.aop.2022.168996} {doi:10.1016/j.aop.2022.168996}. \href{https://arxiv.org/abs/2204.11313} {arXiv:2204.11313}


\end{thebibliography}
\end{document}